\documentclass[11pt]{article}

\usepackage{a4,amsfonts,color,morefloats,pslatex}
\usepackage{amssymb, amsmath,amsthm,latexsym}

\newtheorem{theorem}{Theorem}
\newtheorem{lemma}[theorem]{Lemma}

\newtheorem{corollary}[theorem]{Corollary}
\newtheorem{remark}[theorem]{Remark}

\newtheorem{Open}[theorem]{Open Problem}

\newtheorem{example}[theorem]{Example}

\newcommand{\ord}{{\mathrm{ord}}}

\newcommand{\tr}{{\mathrm{Tr}}}

\newcommand{\gf}{{\mathrm{GF}}}

\newcommand{\Z}{\mathbb{{Z}}}
\newcommand{\ls}{{\mathbb{L}}}

\newcommand{\m}{\mathbb{M}}

\newcommand{\cB}{{\mathcal{B}}}

\newcommand{\C}{{\mathcal{C}}}
\newcommand{\calC}{{\mathcal{C}}}

\newcommand{\N}{\mathbb {N}}

\begin{document}

\title{A Sequence Construction of Cyclic Codes over Finite Fields\label{ra_ch1}}

\author{Cunsheng Ding \\ 
Department of Computer Science and Engineering\\
The Hong Kong University of Science and Technology \\ 
Clear Water Bay, Kowloon, Hong Kong, China \\ 
cding@ust.hk}

\maketitle

\begin{abstract} 
Cyclic codes over finite fields are widely implemented in data storage systems, communication 
systems, and consumer electronics, as they have very efficient encoding and decoding algorithms. 
They are also important in theory, as they are 
closely connected to several areas in mathematics.  There are a few fundamental ways of constructing 
all cyclic codes over finite fields, including the generator matrix approach, the generator 
polynomial approach, the generating idempotent approach, and the $q$-polynomial approach. 
Another one is a sequence approach, which has been intensively investigated in the past decade. 
The objective of this paper is to survey the progress in the past decade  in this direction.  
\end{abstract}


\tableofcontents

\section{Introduction}

Let $q$ be a power of a prime $p$. 
An $[n,k, d]$ code over $\gf(q)$ is a $k$-dimensional subspace of $\gf(q)^n$ 
with minimum (Hamming) nonzero weight $d$. 
Let $A_i$ denote the number of codewords with Hamming weight $i$ in a linear code
$\C$ of length $n$. The {\em weight enumerator} of $\C$ is defined by
$$
1+A_1z+A_2z^2+ \cdots + A_nz^n.
$$
The {\em weight distribution} of $\C$ is the sequence $(1,A_1,\ldots,A_n)$.

An $[n, k, d]$ code over $\gf(q)$ is called \textit{optimal} if there is no 
$[n, k, d']$ code with $d'>d$ or $[n, k', d]$ code with $k'>k$ or $[n', k, d]$ 
code with $n' <n$ over $\gf(q)$ or the parameters $[n, k, d]$ meet a 
bound for linear codes over $\gf(q)$. An 
$[n, k, d]$ code is said to be \textit{almost optimal} if a linear code with 
parameters $[n, k+1, d]$ or $[n, k, d+1]$ or $[n-1, k, d]$ is optimal.   

A vector $(c_0, c_1, \cdots, c_{n-1}) \in \gf(q)^n$ is said to be {\em even-like} 
if $\sum_{i=0}^{n-1} c_i =0$, and is {\em odd-like} otherwise. The {\em even-like subcode} of a 
linear code consists of all the even-like codewords of this linear code.

An $[n,k]$ code over a finite field is said to be {\em cyclic} if 
$(c_0,c_1, \cdots, c_{n-1}) \in \C$ implies $(c_{n-1}, c_0, c_1, \cdots, c_{n-2}) 
\in \C$.  
Let $\gcd(n, q)=1$. We identify a vector $(c_0,c_1, \cdots, c_{n-1}) \in \gf(q)^n$ 
with the polynomial  
$$ 
c_0+c_1x+c_2x^2+ \cdots + c_{n-1}x^{n-1} \in \gf(q)[x]/(x^n-1).  
$$
Then 
a code $\C$ of length $n$ over $\gf(q)$ corresponds to a subset of the quotient ring $\gf(q)[x]/(x^n-1)$. 
The linear code $\C$ is cyclic if and only if the corresponding subset in $\gf(q)[x]/(x^n-1)$ 
is an ideal of the ring $\gf(q)[x]/(x^n-1)$. 
It is well known that every ideal of $\gf(q)[x]/(x^n-1)$ is principal. Let $\C=\langle g(x) \rangle$ be a 
cyclic code, where $g$ is monic and has the least degree. Then $g(x)$ is called the {\em generator polynomial} and 
$h(x)=(x^n-1)/g(x)$ is referred to as the {\em check} polynomial of 
$\C$.  The \emph{dual code}, denoted by $\C^\perp$, of $\C$ has generator polynomial $\bar{h}(x)$, 
which is the reciprocal of $h(x)$. The \emph{complement code}, denoted by $\C^c$, is generated 
by $h(x)$. It is known that $\C^\perp$ and $\C^c$ of $\C$ have the same weight distribution.   

The error correcting capability of cyclic codes may not be as good as some other linear 
codes in general. However, cyclic codes have wide applications in storage and communication 
systems because they have very efficient encoding and decoding algorithms 
\cite{Chie,Forn,Pran}.

Cyclic codes have been studied for decades and a lot of  progress has been made 
\cite{Char,HPbook}.  The total number of cyclic codes 
over $\gf(q)$ and their constructions are closely related to cyclotomic cosets 
modulo $n$, and thus many areas of number theory. Cyclic codes are very important 
in practice, as some cyclic codes are widely implemented in data storage systems and 
communication systems. 

A construction method is said to be \emph{fundamental}, if every cyclic code over any finite 
field can be constructed with this method. Fundamental construction methods of cyclic codes 
include the following: 
\begin{itemize} 
\item The generator matrix (equivalently check matrix) approach. 
\item The generator polynomial (equivalently the check polynomial) approach. 
\item The generating idempotent approach. 
\item The $q$-polynomial approach \cite{DingLing}.  
\end{itemize}  
Another way of   
constructing cyclic codes over $\gf(q)$ with length $n$ is  to use the generator polynomial 
\begin{eqnarray}\label{eqn-defseqcode}
\frac{x^n-1}{\gcd(S(x), x^n-1)}
\end{eqnarray}
where 
$$ 
S(x)=\sum_{i=0}^{n-1} s_i x^i  \in \gf(q)[x]   
$$
and $s^{\infty}=(s_i)_{i=0}^{\infty}$ is a sequence of period $n$ over $\gf(q)$. 
Throughout this paper, we call the cyclic code $\C_s$ with the generator polynomial 
of (\ref{eqn-defseqcode}) the {\em code defined by the sequence} $s^{\infty}$, 
and the sequence $s^{\infty}$ the {\em defining sequence} of the cyclic code $\C_s$. 

It is straightforward to see that every cyclic code of length $n$ over $\gf(q)$ can be 
expressed as $\C_s$ for some sequence $s^\infty$ of period $n$ over $\gf(q)$. 
Because of this, this sequence construction of cyclic codes is \emph{fundamental}. 
An impressive progress in the construction of cyclic codes with this approach has 
been made in the past decade \cite{Ding120,Ding121,Ding13,DingBK15,DZ2014,TQX,Weld}.  

The objective of this paper is to give a survey of recent development in this sequence 
construction of cyclic codes over finite fields. In view that this topic is huge, we 
have to do a selective survey. Our idea is that this paper complements the 
monograph \cite{DingBK15}, so that the two references together could give a well 
rounded treatment of the sequence construction of cyclic codes over finite fields.     
It is hoped that this paper could stimulate further investigations into this sequence 
approach. This paper is an extended version of the survey paper \cite{DingSurvey}.   

\section{Preliminaries} 

In this section, we present basic notation and results of $q$-cyclotomic cosets modulo $n$, 
planar and almost perfect nonlinear functions, 
and sequences that will be employed in subsequent sections.     

\subsection{Some notation fixed throughout this paper}\label{sec-notations} 

Throughout this paper, we adopt the following notation unless otherwise stated: 
\begin{itemize} 
\item $p$ is a prime, $q$ is a positive power of $p$, and $m$ is a positive integer.  
\item $\Z_n=\{0,1,\cdots, n-1\}$, the ring of integers modulo $n$. 
\item $\N_q(x)$ is a function defined by $\N_q(x)=0$ if $x \equiv 0 \pmod{q}$ and $\N_q(x)=1$ otherwise.  
\item $\alpha$ is a generator of $\gf(q^m)^*$, the multiplicative group of $\gf(q^m)$.  
\item $m_a(x)$ is the minimal polynomial of $a \in \gf(q^m)$ over $\gf(q)$.
\item $\tr(x)$ is the trace function from $\gf(q^m)$ to $\gf(q)$.   
\item $\delta(x)$ is a function on $\gf(q^m)$ defined by $\delta(x) =0$ if $\tr(x)=0$ and $\delta(x) =1$ otherwise.   
\item $C_i$ denotes the $q$-cyclotomic coset modulo $n$ containing $i$. 
\item $\Gamma$ is the set of all coset leaders of the $q$-cyclotomic cosets modulo $n$. 
\item For any polynomial $g(x) \in \gf(q)[x]$ with $g(0) \ne 0$, $\bar{g}(x)$ denotes the reciprocal of 
         $g(x)$. 
\item For a cyclic code $\C$ of length $n$ over $\gf(q)$ with generator polynomial $g(x)$, 
$\C^c$ denotes its complement code that is generated by 
      $h(x):=(x^n-1)/g(x)$, and $\C^\perp$ denotes its dual code 
      with generator polynomial $\bar{h}(x)$, i.e., the reciprocal of $h(x)$.       
\item By the Database we mean the collection of tables of best linear codes known maintained by 
         Markus Grassl at http://www.codetables.de/.                   
\end{itemize} 

\subsection{Planar and APN polynomials} 

A function $f: \gf(q^m) \to \gf(q^m)$ is called {\em almost perfect
nonlinear (APN)} if
$$
\max_{a \in \gf(q^m)^*} \max_{b \in \gf(q^m)} |\{x \in \gf(q^m): f(x+a)-f(x)=b\}| =2,
$$
and is referred to as
{\em perfect
nonlinear or planar} if
$$
\max_{a \in \gf(q^m)^*} \max_{b \in \gf(q^m)} |\{x \in \gf(q^m): f(x+a)-f(x)=b\}| =1.
$$

There is no perfect nonlinear (planar) function on $\gf(q^m)$ for even $q$. But there are APN
functions on $\gf(2^m)$. Both planar and APN functions over $\gf(q^m)$ for odd $q$ exist. 
For example, $x^2$ is a planar function on $\gf(q^m)$ for odd $q$, and $x^{q^{m}-2}$ is an APN 
function over $\gf(q^m)$. Perfect nonlinear and APN functions were used to construct linear codes 
in different ways in the literature. Some planar and APN monomials will be employed to construct 
sequences first, and the obtained sequences will be used to construct cyclic codes in subsequent 
sections.

\subsection{The $q$-cyclotomic cosets modulo $n$}\label{sec-cpsets}

Let $\gcd(n, q)=1$. 
The $q$-cyclotomic coset containing $j$ modulo $n$ is defined by 
$$ 
C_j=\{j, qj, q^2j, \cdots, q^{\ell_j-1}j\} \bmod{n} \subset \Z_n
$$
where $\ell_j$ is the smallest positive integer such that $q^{\ell_j}j \equiv j \pmod{n}$, 
and is called the size of $C_j$. It is known that $\ell_j$ divides $m$. The smallest integer 
in $C_j$ is called the {\em coset leader} of $C_j$. Let $\Gamma$ denote the set of all 
coset leaders. By definition, we have 
$$ 
\bigcup_{j \in \Gamma} C_j =\Z_n.  
$$ 

It is well known that $\prod_{j \in C_i} (x-\alpha^j)$ is an irreducible polynomial of degree 
$\ell_i$ over $\gf(q)$ and is the minimal polynomial of $\alpha^i$ over $\gf(q)$. Furthermore,  
the canonical factorization of $x^n-1$ over $\gf(q)$ is given by   
$$ 
x^n-1=\prod_{i \in \Gamma} \prod_{j \in C_i} (x-\alpha^j).  
$$

\subsection{The linear span and minimal polynomial of sequences}

Let $s^L=s_0s_1\cdots s_{L-1}$ be a sequence over $\gf(q)$. The {\em linear 
span} (also called {\em linear complexity}) of $s^L$ is defined to be the smallest positive 
integer $\ell$ such that there are constants $c_0=1, c_1, \cdots, c_\ell \in \gf(q)$ 
satisfying 
\begin{eqnarray*} 
-c_0s_i=c_1s_{i-1}+c_2s_{i-2}+\cdots +c_ls_{i-\ell} \mbox{ for all } \ell \leq i<L. 
\end{eqnarray*} 
In engineering terms, such a polynomial $c(x)=c_0+c_1x+\cdots +c_lx^l$ 
is called the {\em feedback  polynomial} of a shortest linear feedback 
shift register 
(LFSR) that generates $s^L$. Such an integer always exists for finite sequences  $s^L$. When 
$L$ is $\infty$, a sequence $s^{\infty}$ is called a semi-infinite 
sequence. If there is no such an integer for a semi-infinite sequence 
$s^{\infty}$, its linear span is defined to be $\infty$. The linear 
span of the zero sequence is defined to be zero. 
For ultimately periodic semi-infinite sequences such an $\ell$ always 
exists. 

Let $s^{\infty}$ be a sequence of period $L$ over $\gf(q)$. 
Any feedback polynomial of $s^{\infty}$ is called a {\em characteristic 
polynomial}. The characteristic polynomial with the smallest degree is 
called the {\em minimal polynomial} of the periodic sequence $s^{\infty}$. 
Since we require that the constant term of any characteristic polynomial 
be 1, the minimal polynomial of any periodic sequence $s^{\infty}$ must 
be unique. In addition, any characteristic polynomial must be a multiple 
of the minimal polynomial.    

For periodic sequences, there are a few ways to determine their linear
span and minimal polynomials. One of them is given in the following
lemma \cite[p.87, Theorem 5.3]{DXS}.

\begin{lemma}\label{lem-ls1}
Let $s^{\infty}$ be a sequence of period $L$ over $\gf(q)$.
Define
$
S^{L}(x)= \sum_{i=0}^{L-1} s_i x^i  \in \gf(q)[x].
$
Then the minimal polynomial $\m_s(x)$ of $s^{\infty}$ is given by
      \begin{eqnarray}\label{eqn-base1}
      \frac{x^{L}-1}{\gcd(x^{L}-1, S^{L}(x))}
      \end{eqnarray}
and the linear span $\ls_s$ of $s^{\infty}$ is given by
      $
       L-\deg(\gcd(x^{L}-1, S^{L}(x))).
      $
\end{lemma}

The other one is given in the following lemma (\cite[Theorem 3]{Antweiler}, \cite{HK98}). 

\begin{lemma} \label{lem-ls2}
Any sequence $s^{\infty}$ over $\gf(q)$ of period $q^m-1$ has a unique expansion of the form
\begin{equation*}
s_t=\sum_{i=0}^{q^m-2}c_{i}\alpha^{it}, \mbox{ for all } t\ge 0,
\end{equation*}
where $\alpha$ is a generator of $\gf(q^m)^*$ and $c_i \in \gf(q^m)$.
Let the index set be $I=\{i \left.\right| c_i\neq 0\}$, then the minimal polynomial $\m_s(x)$ of $s^{\infty}$ is
$\m_s(x)=\prod_{i\in I}(1-\alpha^i x),$
and the linear span of $s^{\infty}$ is $|I|$.
\end{lemma}

It should be noticed that in some references the reciprocal of $\m_s(x)$ is called the minimal polynomial
of the sequence $s^\infty$. So Lemma \ref{lem-ls2} is a modified version of the original one in \cite{HK98}.

\section{Cyclic codes from combinatorial sequences} 

\subsection{The classical cyclic code $\C_{\gf(q)}(D)$ of a subset $D \in \Z_n$}\label{sec-classicalCodeD} 

Let $n$ be a positive integer, and let $D$ be a subset of $\Z_n$. Define 
$
B_i=D+i
$ 
for all $i \in \Z_n$. Then the pair $(\Z_n, \cB)$ is called an \emph{incidence structure}, 
where $\cB=\{B_0, B_1, \cdots, B_{n-1}\}$. The \emph{incidence matrix} $M_D=(m_{ij})$ of this 
incidence structure is an $n \times n$ matrix, where $m_{ij}=1$ if $j \in B_i$ and $m_{ij}=0$ 
otherwise. By definition, $M_D=(m_{ij})$ is a binary matrix. When $M_D=(m_{ij})$ is viewed as a 
matrix over $\gf(q)$, its row vectors span a cyclic code of length $n$ over $\gf(q)$, which 
is denoted by $\C_{\gf(q)}(D)$ and called the \emph{classical code} of $D$. It is easily seen 
that the generator polynomial of $\C_{\gf(q)}(D)$ is given by  
\begin{eqnarray}\label{eqn-classicalCodeGenerator}
\gcd\left(x^n-1, \, \, \sum_{i \in D} x^i\right), 
\end{eqnarray}
where the greatest common divisor is computed over $\gf(q)$ \cite{DingBK15}[p. 66].

When $D$ has certain combinatorial structures, the cyclic code $\C_{\gf(q)}(D)$ has been well 
studied in the literature \cite{AssmusKey92,DingBK15}. This code is closely related 
to a code dealt with in the next section.

\subsection{The cyclic code of the characteristic sequence of a subset $D \in \Z_n$}

Let $D$ be a subset of $\Z_n$. The {\it characteristic sequence} $s(D)^\infty$ of $D$ is given by
$$ 
s(D)_i=\left\{ \begin{array}{ll} 
            1 & \mbox{ if } i \bmod n \in D, \\
            0 & \mbox{ otherwise.} 
\end{array}
\right. 
$$ 
The binary sequence $s(D)^\infty$ can be viewed as a sequence of period $n$ over any field $\gf(q)$, 
and can be employed to construct the code $\C_{s(D)}$ over $\gf(q)$. For any given pair of $n$ and $q$ with $\gcd(q, n)=1$, 
the subset $D$ must 
be chosen properly, in order to construct a cyclic code $\C_{s(D)}$ with desirable parameters. Intuitively, 
a good choice may be to select a subset $D$ of $\Z_n$ with certain combinatorial structures. It follows from 
the discussions in Section \ref{sec-classicalCodeD} that 
\begin{eqnarray}\label{eqn-codeRelation}
\C_{s(D)}=\C_{\gf(q)}(D)^c. 
\end{eqnarray} 
Hence, in the case that the sequence $s^{\infty}$ over $\gf(q)$ has only entries $0$ and $1$, the 
sequence code $\C_{s(D)}$ is the complement code of the classical code of its support set. This is 
a major connection between the classical construction of cyclic codes with incidence structures and the 
sequence construction of this paper in the special case. However, the two approaches do not include 
each other.  

Let $D$ be a $\kappa$-subset of 
$\Z_n$. The set $D$ is an $(n, \kappa, \lambda)$ {\it difference set} in $(\Z_n, +)$ 
if the multiset 
$$ 
\{ x-y| x, y \in D \}
$$
contains every nonzero element of $\Z_n$ exactly $\lambda$ times. 

Let $D$ be a $\kappa$-subset of 
$\Z_n$. The set $D$ is an $(n, \kappa, \lambda, t)$ {\it almost difference set (ADS)} in $(\Z_n, +)$ 
if the multiset 
$$ 
\{ x-y| x, y \in D \}
$$
contains $t$ nonzero elements of $\Z_n$ exactly $\lambda$ times each and the remaining $n-1-t$ nonzero elements 
$\lambda+1$ times each.  

\begin{example}\label{exam-731ds} 
The Singer difference set in $(\Z_n, +)$ is given by   
$ 
D=\log_{\alpha}\{x \in \gf(2^m): \tr(x)=1\} \subset \Z_n, 
$ 
and has parameters $(2^m-1, 2^{m-1}, 2^{m-2})$, where $\alpha$ is 
generator of $\gf(2^m)^*$ and $n=2^m-1$. 

Its characteristic sequence $(s_t)_{t=0}^\infty$, where $s_t=\tr(\alpha^t)$ for any $t \ge 0$, 
a {\it maximum-length sequence} of period $2^m-1$.  
The minimal polynomial of the Singer sequence is 
equal to the minimal polynomial $m_{\alpha^{-1}}(x)$ of $\alpha^{-1}$ over $\gf(2)$, and 
its linear span is $m$. 
The cyclic code $\C_{s(D)}$ defined by the characteristic sequence of the Singer sequence is 
equivalent to the Hamming code with parameters $[2^m-1, 2^m-1-m, 3]$ and 
has generator polynomial $m_{\alpha^{-1}}(x)$. The code is {\it optimal (perfect)}. 
\end{example} 

A proof of the following results can be found in \cite{DingBK15}[p. 193]. 

\begin{example} 
Let $q=p^s$ be a prime power, where $p$ is a prime, and $s$ is a positive integer, 
and let $m \geq 3$ be a positive integer. Let $\alpha$ be a generator of $\gf(q^m)^*$. 
Put $n=(q^m-1)/(q-1)$. Recall that  
\begin{eqnarray*}
D = \{0 \leq i <n: \tr_{q^m/q} (\alpha^i)=0\} \subset \Z_n 
\end{eqnarray*}
is the Singer difference set in $(\Z_n,+)$ with parameters
\begin{eqnarray*}
\left( \frac{q^m-1}{q-1},  \frac{q^{m-1}-1}{q-1}, \frac{q^{m-2}-1}{q-1}  \right). 
\end{eqnarray*}
Let $s(D)^\infty$ be the characteristic sequence of $D$. Then the cyclic code $\C_{s(D)}^c$ has 
parameters  
\begin{eqnarray}\label{eqn-SingerCode}
\left[ \frac{q^m-1}{q-1}, \binom{p+m-2}{m-1}^s+1, \frac{q^{m-1}-1}{q-1}\right].  
\end{eqnarray}
\end{example} 

Although the parameters of $\C_{s(D)}^c$ are known, the following problem is still open. 

\begin{Open} 
Determine the minimum distance of the code $\C_{s(D)}$ from the characteristic sequence of 
the Singer difference set $D$.  
\end{Open} 

There are many families of difference sets and almost difference sets $D$ in $(\Z_n, +)$. 
In many cases, the dimension and generator polynomial of the classical code $\C_{\gf(q)}(D)$ 
or its complement $\C_{\gf(q)}(D)^c$ (hence, $\C_{\gf(q)}(D)^\perp$) are known. However, their 
minimum distances are open in general. Because of the relation in (\ref{eqn-codeRelation}), 
the dimension and generator polynomial of the 
sequence code $\C_{s(D)}$ are known in many cases, but its minimum distance is known only in some 
special cases. As this is a huge topic with a lot of results, it would be infeasible to survey 
the developments here. Thus, we refer the reader to the monograph \cite{DingBK15} 
for detailed information.   

Cyclotomic classes were employed to define binary sequences, which can be viewed as sequences 
over $\gf(q)$ for any prime power $q$. Such sequences give cyclic codes $\C_{s(D)}$ over $\gf(q)$. The reader 
is referred to \cite{Ding120,Ding13,DingBK15} for detailed information.

\section{Cyclic codes from a construction of sequences from polynomials over $\gf(q^m)$}\label{sec-polycode} 

Given a polynomial $f(x)$ on $\gf(q^m)$, we define its associated sequence 
$s^\infty$ by 
\begin{eqnarray}\label{eqn-sequence}
s_i=\tr(f(\alpha^i+1)) 
\end{eqnarray}
for all $i \ge 0$, where $\alpha$ is a generator of $\gf(q^m)^*$ and $\tr(x)$ denotes 
the trace function from $\gf(q^m)$ to $\gf(q)$. The code $\C_s$ defined by the sequence 
$s^\infty$ in (\ref{eqn-sequence}) is called the \emph{code from the polynomial} $f(x)$ 
for simplicity. 

It was demonstrated in \cite{Ding13,DZ2014,TQX} that the code $\C_s$ may have interesting 
parameters if the polynomial $f$ is properly chosen. The objective of this section is to 
survey cyclic codes $\C_s$ defined by special polynomials $f$ over $\gf(q^m)$. 

\subsection{Cyclic codes from special monomials}\label{sec-Monomialcode}

The following is a list of monomials over $\gf(q^m)$ with good nonlinearity (see \cite[Section 1.7]{DingBK15} 
for definition and details). 
\begin{itemize}
\item $f(x)=x^{q^m-2}$ over $\gf(q^m)$ (APN). 
\item $f(x)=x^{q^k+1}$ over $\gf(q^m)$, where $m/\gcd(m,k)$ and $q$ are odd (planar). 
\item $f(x)=x^{(q^h-1)/(q-1)}$ over $\gf(q^m)$.
\item $f(x)= x^{(3^h+1)/2}$ over $\gf(3^m)$ (planar when $\gcd(h,m)=1$).
\item $f(x)=x^{2^t+3}$ over $\gf(2^m)$ (APN). 
\item $f(x)=x^{e}$ over $\gf(2^m)$, $e=2^{(m-1)/2}+2^{(m-1)/4}-1$ 
and $m \equiv 1 \pmod{4}$ (APN).  
\end{itemize}
When they are plugged into (\ref{eqn-sequence}), sequences over $\gf(q)$ with certain  
properties are obtained. The corresponding sequence codes have interesting parameters. 
The objective of this section is to introduce the parameters of these cyclic codes.  

Let $t$ be a positive integer. We define $T=2^t-1$. For any
odd $a \in \{1,2,3,\cdots,T\}$, define
\begin{equation*}
\epsilon_a^{(t)} =\left\{
\begin{array}{ll}
1, &\textrm{if~} a=2^h-1, \\
\left\lceil {\log_2{T\over a}}\right\rceil \bmod 2,&\textrm{if~}   1\leq a<2^h-1 
\end{array} \right.\ \
\end{equation*}
and
\begin{equation}\label{eqn-def-kappa}
\kappa_a^{(t)} = \epsilon_a^{(t)}~\bmod 2.
\end{equation}
This function $\kappa_a^{(t)}$ will be employed later. 

\subsubsection{Binary cyclic codes from $f(x)=x^{2^{t}+3}$}\label{sec-Welch} 

The monomial $f(x)=x^{2^{t}+3}$ is APN over $\gf(2^{2t+1})$. Both the sequence 
in (\ref{eqn-sequence}) defined by this monomial and the code $\C_s$ are interesting. 

\begin{theorem}\label{thm-Welch} \cite{DZ2014} 
Let $m =2t+1 \ge 7$.
Let $s^{\infty}$ be the sequence of (\ref{eqn-sequence}), where $f(x)=x^{2^{t}+3}$.
Then the linear span $\ls_s$ of $s^{\infty}$ is equal to $5m+1$ and the minimal polynomial $\m_s(x)$
of  $s^{\infty}$ is given by
\begin{eqnarray}\label{eqn-Welch}
\m_s(x)=  (x-1) m_{\alpha^{-1}}(x) m_{\alpha^{-3}}(x) m_{\alpha^{-(2^t+1)}}(x)m_{\alpha^{-(2^t+2)}}(x) m_{\alpha^{-(2^t+3)}}(x).
\end{eqnarray}
The binary code $\C_{s}$ has parameters
$[2^m-1, 2^{m}-2-5m, d]$ and  generator polynomial $\m_s(x)$ of (\ref{eqn-Welch}), where $d \ge 8$.
\end{theorem}

\begin{example}
Let $m=5$ and $\alpha$ be a generator of $\gf(2^m)^*$ with $\alpha^5 + \alpha^2 + 1=0$. Then
the generator polynomial of the code $\C_s$ is
$
\m_s(x)=x^{16} + x^{15} + x^{13} + x^{12} + x^8 + x^6 + x^3 + 1,
$
and $\C_s$ is a $[31, 15, 8]$ binary cyclic code. Its dual is a $[31,16,7]$ cyclic code. Both codes are optimal
according to the Database.
\end{example}

\begin{example}\label{ex-welch2}
Let $m=7$ and $\alpha$ be a generator of $\gf(2^m)^*$ with $\alpha^7 + \alpha + 1=0$. Then
the generator polynomial of the code $\C_s$ is
$
\m_s(x) =  x^{36} + x^{34} + x^{33} + x^{32} + x^{29} + x^{28} + x^{27} +
  x^{26} + x^{25} +
    x^{24} +  x^{21} + x^{12} + x^{11} + x^9 + x^7 + x^6 + x^5 + x^3 + x + 1
$
and $\C_s$ is a $[127, 91, 8]$ binary cyclic code.
\end{example}

It can be seen from Example \ref{ex-welch2}
that the bound on the minimal distance of $\C_s$ in
Theorem \ref{thm-Welch} is tight in certain cases. 
The code $\C_{s}$ in Theorem \ref{thm-Welch} could be optimal in some cases \cite{DZ2014}. 
It would be interesting to settle the following problem.  

\begin{Open} 
Determine the minimum distance of the code $\C_{s}$ in Theorem \ref{thm-Welch}. 
\end{Open} 

\subsubsection{Binary cyclic codes from $f(x)=x^{2^h-1}$}\label{sec-2hminus1}

Consider the monomial $f(x)=x^{2^h-1}$ over $\gf(2^m)$, where $h$ is a positive integer with $1\leq h\leq {\lceil {m\over 2} \rceil}$. As will be demonstrated below, it gives a binary sequence and binary code with 
special parameters. 

\begin{theorem}\label{thm-22mm1}  \cite{DZ2014}
Let $s^{\infty}$ be the sequence of (\ref{eqn-sequence}), where $f(x)=x^{2^h-1}$, $2\leq h\leq {\lceil {m\over 2} \rceil}$. Then the linear span $\ls_s$ of $s^{\infty}$ is given by
\begin{eqnarray}\label{eqn-22m0}
\ls_s =\left\{ \begin{array}{l}
                   \frac{m(2^h+(-1)^{h-1})}{3},~ \mbox{ if $m$ is even,} \\
                   \frac{m(2^h+(-1)^{h-1}) +3}{3},~ \mbox{ if $m$ is odd.}
\end{array}
\right.
\end{eqnarray}
The minimal polynomial 
\begin{equation}\label{eqn-2m31}
\m_s(x) =
 (x-1)^{\N_2(m)} \prod_{1 \le 2j+1 \le 2^h-1 \atop \kappa_{2j+1}^{(h)} =1} m_{\alpha^{-(2j+1)}}(x), 
\end{equation}
where $\kappa_{2j+1}^{(h)}$ was defined in (\ref{eqn-def-kappa}), $\N_2(i) =0$ if $i \equiv 0 \pmod{2}$ and $\N_2(i) =1$ otherwise. 

Let $h \ge 2$.
Then the binary code $\C_{s}$ has parameters
$[2^m-1, 2^{m}-1-\ls_s, d]$ and generator polynomial $\m_s(x)$,
where 
 \begin{eqnarray*}
 d \ge \left\{ \begin{array}{l}
                     2^{h-2}+2 \mbox{ if $m$ is odd and $h>2$,} \\
                     2^{h-2}+1.
                     \end{array}
 \right.
 \end{eqnarray*}
\end{theorem} 

\begin{example}
Let $(m,h)=(7,2)$ and $\alpha$ be a generator of $\gf(2^m)^*$ with $\alpha^7 + \alpha + 1=0$. Then
the generator polynomial of the code $\C_s$ is
$
\m_s(x) = x^8 + x^6 + x^5 + x^4 + x^3 + x^2 + x + 1,
$
and $\C_s$ is a $[127, 119, 4]$ binary cyclic code and optimal according to the Database.
\end{example}

\begin{example}
Let $(m, h)=(7,3)$ and $\alpha$ be a generator of $\gf(2^m)^*$ with $\alpha^7 + \alpha + 1=0$. Then
the generator polynomial of the code $\C_s$ is
$
\m_s(x) = x^{22} + x^{21} + x^{20} + x^{18} + x^{17} + x^{16} + x^{14} +
 x^{13} + x^8 + x^7 + x^6 + x^5 + x^4 + 1
$
and $\C_s$ is a $[127, 105, d]$ binary cyclic code, where $4 \le d \le 8$.
\end{example}

\begin{remark}
The code $\C_s$ of Theorem \ref{thm-22mm1} may be bad when $\gcd(h, m) \ne 1$. In this case the monomial
$f(x)=x^{2^h-1}$ is not a permutation of $\gf(2^m)$. For example, when $(m, h)=(6,3)$, $\C_s$ is a
$[63, 45, 3]$ binary cyclic code, while the best known linear code in the Database has parameters $[63, 45, 8]$.  Hence,
we are interested in this code only for the case that $\gcd(h, m)=1$, which guarantees that $f(x)=x^{2^h-1}$ is a
permutation of $\gf(2^m)$.
\end{remark}

The code $\C_{s}$ in Theorem \ref{thm-22mm1} could be optimal in some cases \cite{DZ2014}. 
The lower bounds on $d$ given in Theorem \ref{thm-22mm1} are quite tight. Nevertheless, 
it would be nice if the following problem could be solved.   

\begin{Open} 
Determine the minimum distance of the code $\C_{s}$ in Theorem \ref{thm-22mm1}. 
\end{Open} 

Let $f(x)=x^{q^h-1}$ over $\gf(q^m)$ and let $s^{\infty}$ be the sequence of (\ref{eqn-sequence}). Then the code 
$\C_s$ was investigated in \cite{LZLK}, where the results of this section were generalised. The reader is referred to 
\cite{LZLK} for detail. 

\subsubsection{Binary cyclic codes from $f(x)=x^e$, $e=2^{(m-1)/2}+2^{(m-1)/4}-1$ and $m \equiv 1 \pmod{4}$}\label{sec-1Niho}

Let $f(x)=x^e$, where $e=2^{(m-1)/2}+2^{(m-1)/4}-1$ and $m \equiv 1 \pmod{4}$.
It is known that $f(x)$ is a permutation of $\gf(2^m)$ and is APN. 
Properties of the binary sequence and binary code defined by $f(x)=x^e$ are documented in the following theorem.  

\begin{theorem}\label{thm-1N2m1}  \cite{DZ2014}
Let $m \ge 9$ be odd.
Let $s^{\infty}$ be the sequence of (\ref{eqn-sequence}). Then the linear span $\ls_s$ of $s^{\infty}$ is given by
\begin{eqnarray}\label{eqn-1N2m0}
\ls_s =\left\{ \begin{array}{l}
                   \frac{m\left(2^{(m+7)/4}+(-1)^{(m-5)/4}\right) +3}{3}, ~\mbox{ if $m \equiv 1 \pmod{8}$}, \\
                   \frac{m\left(2^{(m+7)/4}+(-1)^{(m-5)/4}-6\right) +3}{3}, ~\mbox{ if $m \equiv 5 \pmod{8}$.}
\end{array}
\right.
\end{eqnarray}
The minimal polynomial 
\begin{equation*}\label{eqn-1N2m21}
\m_s(x) = (x-1) \prod_{i=0}^{2^{\frac{m-1}{4}}-1} m_{\alpha^{-i-2^{\frac{m-1}{2}} }}(x)
\prod_{1 \le 2j+1 \le 2^{\frac{m-1}{4}}-1 \atop \kappa_{2j+1}^{((m-1)/4)} =1} m_{\alpha^{-2j-1}}(x)
\end{equation*}
if $m \equiv 1 \pmod{8}$; and
\begin{equation*}\label{eqn-1N2m31}
\m_s(x) =(x-1) \prod_{i=1}^{2^{\frac{m-1}{4}}-1} m_{\alpha^{-i-2^{\frac{m-1}{2}} }}(x)
\prod_{3 \le 2j+1 \le 2^{\frac{m-1}{4}}-1 \atop \kappa_{2j+1}^{((m-1)/4)} =1} m_{\alpha^{-2j-1}}(x)
\end{equation*}
if $m \equiv 5 \pmod{8}$,
where $\kappa_{2j+1}^{(h)}$ was
defined in (\ref{eqn-def-kappa}).

The binary code $\C_{s}$ has parameters
$[2^m-1, 2^{m}-1-\ls_s, d]$ and generator polynomial $\m_s(x)$,
and the minimum weight $d$ has the following bounds: 
\begin{eqnarray}\label{eqn-niho1b}
d \ge \left\{ \begin{array}{ll}
 2^{(m-1)/4} + 2 & \mbox{if } m \equiv 1 \pmod{8}, \\
 2^{(m-1)/4}       & \mbox{if } m \equiv 5 \pmod{8}.
\end{array}
\right.
\end{eqnarray} 
\end{theorem}

\begin{example}
Let $m=5$ and $\alpha$ be a generator of $\gf(2^m)^*$ with $\alpha^5 + \alpha^2 + 1=0$. Then
the generator polynomial of the code $\C_s$ is
$
\m_s(x)=x^6 + x^3 + x^2 + 1,
$
and $\C_s$ is a $[31, 25, 4]$ binary cyclic code and  optimal according to the Database.

\end{example}

\begin{example}
Let $m=9$ and $\alpha$ be a generator of $\gf(2^m)^*$ with $\alpha^9 + \alpha^4 + 1=0$. Then
the generator polynomial of the code $\C_s$ is
$
\m_s(x) =  x^{46} + x^{45} + x^{41} + x^{40} + x^{39} + x^{36} + x^{35} + x^{33} + x^{28} + x^{27} + x^{26} + x^{25} +  x^{24} + x^{22} + x^{21} + x^{20} + x^{19} + x^{14} + x^{12} + x^7 + x^4 +
    x^2 + x + 1
$
and $\C_s$ is a $[511, 465, d]$ binary cyclic code, where $d \ge 6$. The actual minimum weight
may be larger than 6.

\end{example}

The code $\C_{s}$ in Theorem \ref{thm-1N2m1} could be optimal in some cases \cite{DZ2014}. 
The lower bounds on $d$ given in Theorem \ref{thm-22mm1} are reasonably good. It would be 
interesting to work on the following problem.    

\begin{Open} 
Determine the minimum distance of the code $\C_{s}$ in Theorem \ref{thm-1N2m1} or improve 
the lower bounds in (\ref{eqn-niho1b}). 
\end{Open} 

\subsubsection{Binary cyclic codes from $f(x)=x^{2^{2h}-2^h+1}$, where $\gcd(m,h)=1$}\label{sec-Kasami}

Define $f(x)=x^e$, where $e=2^{2h}-2^h+1$ and $\gcd(m,h)=1$. It is known that $f$ is APN under these 
conditions. In this section, we restrict $h$ to the following range: 
\begin{eqnarray}\label{eqn-Hcondition}
1 \le h \le \left\{  \begin{array}{l}
                   \frac{m-1}{4} \mbox{ if } m \equiv 1 \pmod{4}, \\
                   \frac{m-3}{4} \mbox{ if } m \equiv 3 \pmod{4}, \\
                   \frac{m-4}{4} \mbox{ if } m \equiv 0 \pmod{4}, \\
                   \frac{m-2}{4} \mbox{ if } m \equiv 2 \pmod{4}.
\end{array}
\right.
\end{eqnarray}

Some parameters of the binary sequence and the code defined by $f(x)=x^e$ are given in 
the following theorem. 

\begin{theorem}\label{thm-K1N2m1} \cite{DZ2014, MSZ22}
Let $h$ satisfy the conditions of (\ref{eqn-Hcondition}).
Let $s^{\infty}$ be the sequence of (\ref{eqn-sequence}). Then the linear span $\ls_s$ of $s^{\infty}$ is given by
\begin{eqnarray}\label{eqn-K1N2m0}
\ls_s =\left\{ \begin{array}{l}
                   \frac{m\left(2^{(h+2}+(-1)^{h-1}\right) +3\N_2(m)}{3} \mbox{ if $h$ is even,} \\
                   \frac{m\left(2^{h+2}+(-1)^{h-1}-6\right) +3\N_2(m)}{3} \mbox{ if $h$ is odd.}
\end{array}
\right.
\end{eqnarray}
The minimal polynomial 
\begin{equation*}\label{eqn-K1N2m21}
\m_s(x) = (x-1)^{\N_2(m)} \prod_{i=0}^{2^{h}-1} m_{\alpha^{-i-2^{m-h} }}(x)
\prod_{1 \le 2j+1 \le 2^{h}-1 \atop \kappa_{2j+1}^{h} =1} m_{\alpha^{-2j-1}}(x)
\end{equation*}
if $h$ is even; and
\begin{equation*}\label{eqn-K1N2m31}
\m_s(x) =(x-1)^{\N_2(m)} \prod_{i=1}^{2^{h}-1} m_{\alpha^{-i-2^{m-h} }}(x)
\prod_{3 \le 2j+1 \le 2^{h}-1 \atop \kappa_{2j+1}^{h} =1} m_{\alpha^{-2j-1}}(x)
\end{equation*}
if $h$ is odd,
where  $\kappa_{2j+1}^{(h)}$ was
defined in (\ref{eqn-def-kappa}).

The code $\C_{s}$ has parameters
$[2^m-1, 2^{m}-1-\ls_s, d]$ and generator polynomial $\m_s(x)$,
 and the minimum weight $d$ has the following
bounds:
\begin{eqnarray}\label{eqn-Kniho1b}
d \ge \left\{ \begin{array}{ll}
 2^{h} + 2 & \mbox{if $h$ is even and $m$ is odd,}  \\
 2^{h} + 1 & \mbox{if $h$ is even and $m$ is even,}  \\ 
 2^{h}       & \mbox{if $h$ is odd.}
\end{array}
\right.
\end{eqnarray}
\end{theorem}

\begin{example}
Let $(m,h)=(5,2)$ and $\alpha$ be a generator of $\gf(2^m)^*$ with $\alpha^5 + \alpha^2 + 1=0$. Then
the generator polynomial of the code $\C_s$ is
$
\m_s(x)= x^{16} + x^{14} + x^{10} + x^{9} + x^8 + x^7 + x^5 + x^4 + x^3 + x^2 + x+ 1
$
and $\C_s$ is a $[31, 15, 8]$ binary cyclic code. Its dual is a $[31,16,7]$ cyclic code. Both codes are optimal according to the Database. 
\end{example}

\begin{example}
Let $(m,h)=(7,2)$ and $\alpha$ be a generator of $\gf(2^m)^*$ with $\alpha^7 + \alpha + 1=0$. Then
the generator polynomial of the code $\C_s$ is
$
\m_s(x) =  x^{36} + x^{28} + x^{27} + x^{23} + x^{21} + x^{20} + x^{18} +
  x^{13} + x^{12} + x^9 + x^7 + x^6 + x^5  + 1
$
and $\C_s$ is a $[127, 91, 8]$ binary cyclic code.

\end{example}

The code $\C_{s}$ in Theorem \ref{thm-K1N2m1} could be optimal in some cases \cite{DZ2014}. 
It would be interesting to attack the following two problems.    

\begin{Open} 
Determine the minimum distance of the code $\C_{s}$ in Theorem \ref{thm-K1N2m1}. 
\end{Open}

The dimension and lower bounds on the minimum weight of the code $\C_s$ of this section were developed 
in \cite{MSZ22} when $h$ satisfies
\begin{eqnarray}\label{eqn-Hcondition2}
\left\{  \begin{array}{l}
                 \frac{m-1}{2} \ge h >                  \frac{m-1}{4} \mbox{ if } m \equiv 1 \pmod{4}, \\
                 \frac{m-3}{2} \ge h >                   \frac{m-3}{4} \mbox{ if } m \equiv 3 \pmod{4}, \\
                 \frac{m-4}{2} \ge h >                   \frac{m-4}{4} \mbox{ if } m \equiv 0 \pmod{4}, \\
                 \frac{m-2}{2} \ge h >                   \frac{m-2}{4} \mbox{ if } m \equiv 2 \pmod{4}.
\end{array}
\right.
\end{eqnarray}

\subsubsection{Binary cyclic codes from $f(x)=x^{2^m-2}$ over $\gf(2^m)$}\label{sec-binaryinversecode} 

Let $\rho_i$ denote the total number of even integers in the $2$-cyclotomic coset $C_i$ modulo $2^m-1$. 
We then define 
\begin{eqnarray}\label{eqn-defnu}
\nu_i=\frac{m\rho_i}{\ell_i} \bmod{2} 
\end{eqnarray} 
for each $i \in \Gamma$, where $\ell_i=|C_i|$ and $\Gamma$ denotes the set of coset leaders modulo $n=2^m-1$.

It is known that $f(x)=x^{2^m-2}$ over $\gf(2^m)$ is APN. For the binary sequence and code defined by 
this monomial, we have the following.  

\begin{theorem}\label{thm-InverseCode} \cite{Ding121}
Let $s^{\infty}$ be the sequence of (\ref{eqn-sequence}), where $f(x)=x^{2^m-2}$. Then the linear span $\ls_s$
of $s^{\infty}$ is equal to $(n+1)/2$ and the minimal polynomial $\m_s(x)$ of  $s^{\infty}$ is given by 
\begin{equation}\label{eqn-gpInverse}
\m_s(x)=\prod_{j \in \Gamma, \nu_j=1} m_{\alpha^{-j}}(x).  
\end{equation} 

The binary code $\C_{s}$ has parameters $[2^m-1, 2^{m-1}-1, d]$ and generator polynomial $\m_s(x)$. 
If $m$ is odd, the minimum distance $d$ of $\C_{s}$ is even and satisfies $d^2-d+1 \ge n$, 
and the dual code $\C_{s}^\perp$ has parameters $[2^m-1, 2^{m-1}, d^\perp]$, where $d^\perp$ 
satisfies that  $(d^\perp)^2-d^\perp+1 \ge n$.  
\end{theorem} 

\begin{example}\label{exam-qrc} 
Let $m=5$ and $\alpha$ be a generator of $\gf(2^m)^*$ with $\alpha^5 + \alpha^2 + 1=0$. Then 
the generator polynomial $\m_s(x)$ of the code $\C_s$ is 
\begin{eqnarray*} 
\lefteqn{(x+1) m_{\alpha^{-3}}(x)  m_{\alpha^{-5}}(x) m_{\alpha^{-15}}(x) } \\ 
& = x^{16} + x^{14} + x^{13} + x^{10} + x^9 + x^8 + x^7 + x^6 + x^5 + x^2 + x + 1       
\end{eqnarray*}
and $\C_s$ is a $[31, 15, 8]$ binary cyclic code. Its dual is a $[31,16,7]$ cyclic code. Both codes are optimal 
according to the Database.  
\end{example}

When $f(x)=x^{q^m-2}$ and $q>2$, the dimension of the code $\C_{s}$ over $\gf(q)$ was settled in \cite{TQX}. 
But no lower bound on the minimum distance of $\C_{s}$ is developed. 

\subsubsection{Cyclic codes from $f(x)=x^{q^\kappa+1}$,  
where $m/\gcd(m,\kappa)$ and $q$ are odd}\label{sec-DOGold} 

Let $f(x)=x^{q^\kappa+1}$, where $m/\gcd(m,\kappa)$ and $q$ are odd. 
It is known that $f$ is planar. Properties of the sequence and code 
defined by this monomial are described below. 

\begin{theorem}\label{thm-DOGold} \cite{Ding121}
Let $m$ be odd. 
Let $s^{\infty}$ be the sequence of (\ref{eqn-sequence}), where $f(x)=x^{q^\kappa+1}$. 
Then the linear span $\ls_s$ of $s^{\infty}$ is equal to $2m+\N_p(m)$ and the minimal polynomial $\m_s(x)$ 
of  $s^{\infty}$ is given by 
\begin{equation}\label{eqn-DOGold}
\m_s(x)= (x-1)^{\N_p(m)} m_{\alpha^{-1}}(x) m_{\alpha^{-(p^\kappa+1)}}(x), 
\end{equation} 
where 
$\N_p(i) =0$ if $i \equiv 0 \pmod{p}$ and $\N_p(i) =1$ otherwise. 

The code $\C_{s}$ has parameters 
$[n, n-2m-\N_p(m), d]$ and generator polynomial $\m_s(x)$, where 
\begin{eqnarray*} 
\left\{ \begin{array}{ll}
d =4             & \mbox{ if $q=3$ and $m \equiv 0 \pmod{p}$,}  \\
4 \le d \le 5  & \mbox{ if $q=3$ and $m \not\equiv 0 \pmod{p}$,} \\ 
d=3              & \mbox{ if $q >3$ and $m \equiv 0 \pmod{p}$,} \\ 
3 \le d  \le 4 & \mbox{ if $q >3$ and $m \not\equiv 0 \pmod{p}$.} 
\end{array} 
\right. 
\end{eqnarray*}   
\end{theorem} 

\begin{example} 
Let $(m,\kappa, q)=(3,1, 3)$ and $\alpha$ be a generator of $\gf(r)^*$ with $\alpha^3 + 2\alpha + 1=0$. Then 
$\C_s$ is a $[26, 20, 4]$ ternary code with generator polynomial  
$  
\m_s(x)= x^6 + 2x^5 + 2x^4+ x^3 + x^2 + 2x+ 1.    
$
This cyclic code is an optimal linear code according to the Database.  
\end{example}

\begin{example} 
Let $(m,\kappa, q)=(4,4, 3)$ and $\alpha$ be a generator of $\gf(r)^*$ with $\alpha^4 + 2\alpha^3 + 2=0$. Then 
$\C_s$ is a $[80, 71, 5]$ ternary code with generator polynomial  
$  
\m_s(x)= x^9 + 2x^8 + x^7 + 2x^6 + x^4 + x^2 + 1.    
$
This cyclic code is an optimal linear code according to the Database. 
\end{example}

Extending the work of \cite{YCD}, one can determine the weight distribution 
of $\C_{s}^\perp$. With the MacWilliams identity, one can settle the minimum 
distance of the code $\C_{s}$ in Theorem \ref{thm-DOGold}. 

\subsubsection{Cyclic codes from $f(x)=x^{(q^h-1)/(q-1)}$}\label{sec-qhminus1} 

Let $h$ be a positive integer satisfying the following condition: 
\begin{eqnarray}\label{eqn-qm1cond} 
\begin{array}{l} 
1 \le h \le \left\{ \begin{array}{l} 
                      (m-1)/2 \mbox{ if $m$ is odd,} \\
                      m/2 \mbox{ if $m$ is even.}   
                           \end{array} 
                           \right. 
\end{array} 
\end{eqnarray}

Let $J \ge t \ge 2$, and let $\N(J, t)$ denote the total number of vectors 
$(i_1, i_2, \cdots, i_{t-1})$ with $1 \le i_1<i_2<\cdots < i_{t-1}<J$. 
By definition, we have the following recursive formula: 
\begin{eqnarray}\label{eqn-recur}
\N(J, t)=\sum_{j=t-1}^{J-1} \N(j, t-1). 
\end{eqnarray} 
It is easily seen that  
\begin{eqnarray}\label{eqn-J2}
\N(J,2)=J-1 \mbox{ for all } J \ge 2  
\end{eqnarray} 
and 
\begin{eqnarray}\label{eqn-J3}
\N(J,3)= \frac{(J-1)(J-2)}{2} \mbox{ for all } J \ge 3.   
\end{eqnarray} 
It then follows from (\ref{eqn-recur}),  (\ref{eqn-J2}) and (\ref{eqn-J3}) that 
\begin{eqnarray}\label{eqn-J4}
\N(J, 4) 
= \sum_{j=3}^{J-1} \N(j, 3) 
= \sum_{j=3}^{J-1} \frac{(J-1)(J-2)}{2}  
= \frac{J^3-6J^2+11J-6}{6}. 
\end{eqnarray} 
By definition, we have 
\begin{eqnarray}\label{eqn-JJ}
\N(t,t)=1 \mbox{ for all } t \ge 2.   
\end{eqnarray} 
For convenience, we define $\N(J,1)=1$ for all $J \ge 1$.

\begin{theorem}\label{thm-2mm1} \cite{Ding121}
Let $h$ satisfy the condition of (\ref{eqn-qm1cond}).  
Let $s^{\infty}$ be the sequence of (\ref{eqn-sequence}), where $f(x)=x^{(q^h-1)/(q-1)}$. Then the linear 
span $\ls_s$ and minimal polynomial $\m_s(x)$ of $s^{\infty}$ are given by 
\begin{eqnarray*}\label{eqn-2m0} 
\ls_s = \left(\N_p(h)+ \sum_{t=1}^{h-1} \sum_{u=1}^{h-1} \N_p(h-u) \N(u, t) \right) m + \N_p(m)  
\end{eqnarray*} 
and 
\begin{eqnarray*}\label{eqn-2m21}
\m_s(x) 
&=& (x-1)^{\N_p(m)}  m_{\alpha^{-1}}(x)^{\N_p(h)}  \prod_{1 \le u \le h-1 \atop \N_p(h-u)=1} m_{\alpha^{-(q^0+q^u)}}(x)   
       \times \nonumber \\
& & \prod_{t=2}^{h-1}   \prod_{t \le u \le h-1 \atop \N_p(h-u)=1}  
\prod_{1 \le i_1<\cdots < i_{t-1} < u} m_{\alpha^{-(q^0+\sum_{j=1}^{t-1} q^{i_j}+ q^u)}}(x).              
\end{eqnarray*} 
The code $\C_{s}$ has parameters 
$[n, n-\ls_s, d]$ and generator polynomial $\m_s(x)$. 
\end{theorem} 

\begin{Open} 
Determine the minimum distance of the code $\C_{s}$ in Theorem \ref{thm-2mm1} or develop a tight 
lower bound on it. 
\end{Open} 

As a corollary of Theorem \ref{thm-2mm1}, we have the following.  
\begin{corollary}\label{cor-qh1} 
Let $h =3$. 
The code $\C_{s}$ of Theorem \ref{thm-2mm1} has parameters  
$[n, n-\ls_s, d]$ and generator polynomial $\m_s(x)$ given by 
\begin{eqnarray*}
\m_s(x) = 
(x-1)^{\N_p(m)} m_{\alpha^{-1}}(x) m_{\alpha^{-1-q}}(x) m_{\alpha^{-1-q^2}}(x)  m_{\alpha^{-1-q-q^2}}(x)   
\end{eqnarray*}
if $p \ne 3$, and 
\begin{eqnarray*}
\m_s(x) = 
(x-1)^{\N_p(m)} m_{\alpha^{-1-q}}(x) m_{\alpha^{-1-q^2}}(x)  m_{\alpha^{-1-q-q^2}}(x)   
\end{eqnarray*}
if $p=3$, where 
\begin{eqnarray}
\ls_s=\left\{ \begin{array}{l}
4m+\N_p(m) \mbox{ if } p \ne 3, \\
3m+\N_p(m) \mbox{ if } p = 3. 
\end{array}
\right. 
\end{eqnarray}
In addition,  
\begin{eqnarray*} 
\left\{ \begin{array}{ll}
3 \le d \le 8  & \mbox{if $p=3$ and $\N_p(m)=1$,} \\
3 \le d \le 6  & \mbox{if $p=3$ and $\N_p(m)=0$,} \\
3 \le d \le 8  & \mbox{if $p>3$.}  
\end{array}
\right. 
\end{eqnarray*}
\end{corollary} 

\begin{example} 
Let $(m,h, q)=(4,3,3)$ and $\alpha$ be a generator of $\gf(r)^*$ with $\alpha^4 +2\alpha^3+2=0$. Then 
$\C_s$ is a $[80, 69, 5]$ ternary code with generator polynomial  
$  
\m_s(x)= x^{11} + 2x^8 + 2x^6 + 2x^5 + 2x^4 + x^3 + 2x^2 + x + 2.    
$ 
This is an almost optimal linear code according to the Database.  The known optimal linear code has parameters $[80, 69, 6]$ 
which is not cyclic. Notice that $h >m/2$. Hence, the parameters of this code do not 
agree with those of the code in Corollary \ref{cor-qh1}. 
In this case $f(x)$ is a 
permutation.  
\end{example} 

\begin{example} 
Let $(m,h, q)=(5,3,3)$ and $\alpha$ be a generator of $\gf(r)^*$ with $\alpha^5 +2\alpha+1=0$. Then 
$\C_s$ is a $[242, 226, d]$ ternary code with generator polynomial  
$  
\m_s(x)= x^{16} + 2x^{14} + 2x^{12} + 2x^{11} + x^{10} + x^9 + x^6 + x^3 + 2x^2 + 2.    
$ 
Notice that $h >m/2$. However, the parameters of this code do  
agree with those of the code in Corollary \ref{cor-qh1}. 
In this case $f(x)$ is a 
permutation.  
\end{example} 

\begin{example} 
Let $(m,h, q)=(6,3,3)$ and $\alpha$ be a generator of $\gf(r)^*$ with $\alpha^6 + 2\alpha^4 + \alpha^2 + 2\alpha + 2=0$. Then 
$\C_s$ is a $[728, 710, d]$ ternary code with generator polynomial  
\begin{eqnarray*}
\m_s(x) = x^{18} + 2x^{15} + 2x^{14} + 2x^{13} + 2x^{11} + x^{10} 
 + 2x^9 + x^8 + x^6 + 2x^4 + x^3 + x^2 + 2.    
\end{eqnarray*} 
Notice that $h = m/2$. Hence, the parameters of this code  
agree with those of the code in Corollary \ref{cor-qh1}. 
In this case $f(x)$ is not a 
permutation. In fact, $\gcd((q^h-1)/(q-1), q^m-1)=13$.  
\end{example}

\begin{Open} 
For the code $\C_s$ of Corollary \ref{cor-qh1}, do the following lower bounds hold?  
\begin{eqnarray*} 
d \ge \left\{ \begin{array}{ll}
5   & \mbox{when $p=3$ and $\N_p(m)=1$,} \\
4   & \mbox{when $p=3$ and $\N_p(m)=0$,} \\ 
6   & \mbox{when $p>3$ and $\N_p(m)=1$,} \\
5   & \mbox{when $p>3$ and $\N_p(m)=0$.} 
\end{array}
\right. 
\end{eqnarray*}
\end{Open} 

The work of this section was extended in \cite{RK2017}, where the restriction in (\ref{eqn-qm1cond}) was removed. 

\subsubsection{Cyclic codes from $f(x)=x^{(3^h+1)/2}$}\label{sec-CM} 

Let $h$ be a positive integer satistying the following conditions: 
\begin{eqnarray}\label{eqn-CM2m1cond} 
\left\{ 
\begin{array}{l}  
h \mbox{ is odd,} \\
\gcd(m, h)=1, \\
3 \le h \le \left\{ \begin{array}{l} 
                      (m-1)/2 \mbox{ if $m$ is odd,} \\
                      m/2 \mbox{ if $m$ is even.}   
                           \end{array} 
                           \right. 
\end{array} 
\right. 
\end{eqnarray}

\begin{theorem}\label{thm-CM2mm1} \cite{Ding121} 
Let $h$ satisfy the third condition of (\ref{eqn-CM2m1cond}).  
Let $s^{\infty}$ be the sequence of (\ref{eqn-sequence}), where $f(x)=x^{(3^h+1)/2}$. Then the linear 
span $\ls_s$ and minimal polynomial $\m_s(x)$ of $s^{\infty}$ are given by 
\begin{eqnarray*}\label{eqn-CM2m0} 
\ls_s &=& \N_3(m) + \left(\sum_{i=0}^h\N_3(h-i+1) \right)m + \nonumber \\ 
         & &  \left( \sum_{t=2}^h \N(h,t) + \sum_{t=2}^{h-1} \sum_{i_t=t}^{h-1} \N_3(h-i_t+1) \N(i_t, t) \right) m   
\end{eqnarray*} 
and 
\begin{eqnarray*}\label{eqn-CM2m21}
\m_s(x) 
&=& (x-1)^{ \N_3(m)}  m_{\alpha^{-1}}(x)^{\N_3(h+1)} m_{\alpha^{-2}}(x) 
      \prod_{t=1}^{h-1} \prod_{1 \le i_1 < \cdots < i_t \le h-1} m_{\alpha^{-(2+\sum_{j=1}^t 3^{i_j})}}(x) \times \\  
& & \prod_{1 \le u \le h-1 \atop \N_3(h-u+1)=1} m_{\alpha^{-(1+3^u)}}(x)  
 \prod_{t=2}^{h-1}   \prod_{t \le i_t \le h-1 \atop \N_3(h-i_t+1)=1}   
      \prod_{1 \le i_1<\cdots < i_{t-1} < i_t} m_{\alpha^{-(1+\sum_{j=1}^{t} 3^{i_j})}}(x),            
\end{eqnarray*} 
where $\N_3(j)$ and $\N(j, t)$ were defined in Sections \ref{sec-notations} and \ref{sec-qhminus1} respectively.  

Furthermore, the code $\C_{s}$ has parameters 
$[n, n-\ls_s, d]$ and generator polynomial $\m_s(x)$. 
\end{theorem} 

As shown in Theorem \ref{thm-CM2mm1}, the linear span and the minimal polynomial of the sequence 
$s^\infty$ have a complex formula. It looks difficult to discover further properties of the code 
in Theorem \ref{thm-CM2mm1}.  

\begin{Open} 
Determine the minimum distance of the code $\C_{s}$ in Theorem \ref{thm-CM2mm1} or develop a tight 
lower bound on it. 
\end{Open} 

As a corollary of Theorem \ref{thm-CM2mm1}, we have the following.  
\begin{corollary}\label{cor-CMqh1} 
Let $h =3$. 
The code $\C_{s}$ of Theorem \ref{thm-CM2mm1} has parameters  
$[n, n-\ls_s, d]$ and the  generator polynomial $\m_s(x)$ given by 
\begin{eqnarray*}
\m_s(x) 
= (x-1)^{\N_3(m)} m_{\alpha^{-1}}(x) m_{\alpha^{-2}}(x)  m_{\alpha^{-5}}(x) 
  m_{\alpha^{-10}}(x) m_{\alpha^{-11}}(x)  m_{\alpha^{-13}}(x)  m_{\alpha^{-14}}(x),       
\end{eqnarray*}
where 
$ 
\ls_s=7m+\N_3(m). 
$ 
In addition, 
\begin{eqnarray*} 
\left\{ \begin{array}{ll}
5 \le d \le 16  & \mbox{if $\N_3(m)=1$,} \\
4 \le d \le 16  & \mbox{if $\N_3(m)=0$.} 
\end{array}
\right. 
\end{eqnarray*}
\end{corollary} 

\begin{Open} 
For the code $\C_s$ of Corollary \ref{cor-CMqh1}, do the following lower bounds hold?  
\begin{eqnarray*} 
d \ge \left\{ \begin{array}{ll}
9   & \mbox{when $\N_p(m)=1$,} \\
8   & \mbox{when $\N_p(m)=0$.}  
\end{array}
\right. 
\end{eqnarray*}
\end{Open} 

Let $f(x)=x^{(p^h+1)/2}$ over $\gf(p^m)$ and let $s^{\infty}$ be the sequence of (\ref{eqn-sequence}). 
Then the code $\C_s$ was studied in \cite{RK2017}, where the work of this section was generalised.

\subsubsection{Cyclic codes from $f(x)=x^{q^\ell+2}$ over $\gf(q^m)$}

In this section, we introduce cyclic codes defined by the monomial $f(x)=x^{q^\ell+2}$ over $\gf(q^m)$, 
where $m=2\ell$. 

\begin{lemma}\label{lem-LZLK} \cite{LZLK}
Let $m=2\ell$ with $\ell$ being a positive integer, and let $s^{\infty}$ be the sequence of (\ref{eqn-sequence}) 
defined by the monomial $f(x)=x^{q^\ell+2}$ over $\gf(q^m)$.  

When $q \neq 2$, the linear 
span $\ls_s$ and minimal polynomial $\m_s(x)$ of $s^{\infty}$ are given by 
$$ 
\ls_s = 2m+\N_p(3)m+\N_p(4)\ell+\N_p(m) 
$$
and 
\begin{eqnarray*}
\m_s(x)= 
\left\{ 
\begin{array}{ll}
m_{\alpha^{-1}}(x) m_{\alpha^{-2}}(x) m_{\alpha^{-(q^\ell +2)}}(x), & p=2, \\
(x-1)^{\N_p(m)} m_{\alpha^{-2}}(x) m_{\alpha^{-(q^\ell +1)}}(x) m_{\alpha^{-(q^\ell +2)}}(x), & p=3, \\
(x-1)^{\N_p(m)} m_{\alpha^{-1}}(x)  m_{\alpha^{-2}}(x) m_{\alpha^{-(q^\ell +1)}}(x) m_{\alpha^{-(q^\ell +2)}}(x), & p>3, \\
\end{array}
\right. 
\end{eqnarray*}
where $\alpha$ is a generator of $\gf(q^m)$. 

When $q=2$,  the linear 
span $\ls_s=m$ and minimal polynomial $\m_s(x)$ of $s^{\infty}$ is $\m_s(x)=m_{\alpha^{-(2^{\ell-1} +1)}}(x)$.  

\end{lemma} 

\begin{theorem} \cite{LZLK}
Let $\C_s$ be the cyclic code defined by the sequence $s^\infty$ in Lemma \ref{lem-LZLK}. Then we have the 
following. 
\begin{itemize} 
\item When $q \neq 2$, $\C_s$ has the generator polynomial $\m_s(x)$ defined in Lemma \ref{lem-LZLK} and 
parameters $[q^m-1, q^m-1-(2m+\N_p(3)m+\N_p(4)\ell+\N_p(m)), d]$, where 
\begin{eqnarray*}
\left\{ 
\begin{array}{ll}
3 \leq d \leq 6, & \mbox{ if }  p=2 \mbox{ or } p=3, \\
3 \leq d \leq 8, & \mbox{ if } p>3. 
\end{array}
\right. 
\end{eqnarray*} 
\item When $q=2$, $\C_s$ has the generator polynomial $\m_s(x)$ defined in Lemma \ref{lem-LZLK} and 
parameters $[2^m-1, 2^m-1-m, d]$, where $d=2$ if $\ell$ is even and $d=3$ otherwise.  
\end{itemize}
\end{theorem}

\subsection{Cyclic codes from Dickson polynomials}\label{sec-1stDPcode} 

In this section, we survey known results on cyclic codes from Dickson polynomials over 
finite fields. All the results presented in this subsection come from \cite{DingDickson}. 

\subsubsection{Dickson polynomials over $\gf(q^m)$}\label{sec-DPAPNPN} 

In 1896, Dickson introduced the following family of polynomials over 
$\gf(q^m)$ \cite{Dick96}: 
\begin{eqnarray}\label{eqn-1stDP}
D_h(x, a)=\sum_{i=0}^{\lfloor \frac{h}{2} \rfloor} \frac{h}{h-i} \binom{h-i}{i}  (-a)^i x^{h-2i}, 
\end{eqnarray} 
where $a \in \gf(q^m)$ and $h \ge 0$ is called the {\em order} of the polynomial. This family is 
referred to as the {\em Dickson polynomials of the first kind}.

Dickson polynomials of the second kind over $\gf(q^m)$ are defined by  
\begin{eqnarray}\label{eqn-2ndDP}
E_h(x, a)=\sum_{i=0}^{\lfloor \frac{h}{2} \rfloor} \binom{h}{h-i} (-a)^i x^{h-2i}, 
\end{eqnarray} 
where $a \in \gf(q^m)$ and $h \ge 0$ is called the {\em order} of the polynomial.

Dickson polynomials are an interesting topic of mathematics and engineering, and have many applications. 
For example, the Dickson polynomials $D_5(x, a)=x^5-ux-u^2x$ over $\gf(3^m)$ are 
employed to construct a family of planar functions \cite{CM,DY06}, and those planar 
functions give two families of commutative presemifields, planes,  several classes of 
linear codes \cite{CDY,YCD}, and two families of skew Hadamard difference sets \cite{DY06}. 
The reader is referred to \cite{LMT} for detailed information about Dickson polynomials. 
In subsequent subsections, we survey cyclic codes derived from Dickson polynomials.

\subsubsection{Cyclic codes from the Dickson polynomial $D_{p^u}(x,a)$}\label{sec-tracex}

Since $q$ is a power of $p$, it is known that $D_{hp}(x,a)=D_{h}(x,a)^p$ \cite[Lemma 2.6 ]{LMT}.  
It then follows that 
$ 
D_{p^u}(x,a)=x^{p^u}
$ 
for all $a \in \gf(q^m)$. 

The code $\C_s$ over $\gf(q)$ defined by the Dickson polynomial $f(x)=D_{p^u}(x,a)=x^{p^u}$ 
over $\gf(q^m)$ has the following parameters.     

\begin{theorem}\label{thm-tracex} 
The code $\C_{s}$ defined by the Dickson polynomial  $D_{p^u}(x,a)=x^{p^u}$ has parameters 
$[n, n-m-\delta(1), d]$ and generator polynomial $\m_s(x)=(x-1)^{\delta(1)} m_{\alpha^{-p^u}}(x)$,    
where 
\begin{eqnarray*} 
d=\left\{ \begin{array}{l} 
4 \mbox{ if }  q=2 \mbox{ and } \delta(1)=1, \\
3 \mbox{ if }  q=2 \mbox{ and } \delta(1)=0, \\
3 \mbox{ if }  q>2 \mbox{ and } \delta(1)=1, \\
2 \mbox{ if }  q>2 \mbox{ and } \delta(1)=0,  
 \end{array} 
 \right. 
 \end{eqnarray*} 
and the function $\delta(x)$ and the polynomial $m_{\alpha^j}(x)$ were defined in Section \ref{sec-notations}. 
\end{theorem}

When $q=2$, the code of Theorem \ref{thm-tracex} is equivalent to the binary Hamming weight or 
its even-weight subcode, and is thus optimal. The code is either optimal or almost optimal with 
respect to the Sphere Packing Bound.

\subsubsection{Cyclic codes from $D_2(x,a)=x^2-2a$}\label{} 

In this section we consider the code $\C_s$ defined by  $f(x)=D_2(x,a)=x^2-2a$ 
over $\gf(q^m)$. When $p=2$, this code was treated in Section \ref{sec-tracex}. When 
$p>2$, the following theorem is a variant of Theorem 5.2 in \cite{Ding121}, but 
has much stronger conclusions on the minimum distance of the code.  

\begin{theorem}\label{thm-DP2square} 
Let $p>2$ and $m \geq 3$. 
The code $\C_{s}$ defined by $f(x)=D_2(x,a)=x^2-2a$ has parameters 
$[n, n-2m-\delta(1-2a), d]$ and generator polynomial 
\begin{equation*} 
\m_s(x)= (x-1)^{\delta(1-2a)} m_{\alpha^{-1}}(x) m_{\alpha^{-2}}(x), 
\end{equation*} 
where 
\begin{eqnarray*} 
d=\left\{ \begin{array}{ll} 
4   & \mbox{ if }  q=3 \mbox{ and } \delta(1-2a)=0, \\
5  & \mbox{ if }  q=3 \mbox{ and } \delta(1-2a)=1, \\
3   & \mbox{ if }  q>3 \mbox{ and } \delta(1-2a)=0, \\
4 & \mbox{ if }  q>3 \mbox{ and } \delta(1-2a)=1,  
 \end{array} 
 \right. 
 \end{eqnarray*} 
 and the function $\delta(x)$ and the polynomial $\m_{\alpha^j}(x)$ were defined in Section \ref{sec-notations}. 
\end{theorem}

The code of Theorem \ref{thm-DP2square} is either optimal or almost optimal for all $m \geq 2$.

\subsubsection{Cyclic codes from $D_3(x,a)=x^3-3ax$}\label{sec-order3} 

In this section we treat the code $\C_s$ defined by the Dickson polynomial 
$D_3(x,a)=x^3-3ax$. We need to distinguish among the three cases: $p=2$, 
$p=3$ and $p \ge 5$. The case that $p=3$ was covered in Section \ref{sec-tracex}. 
So we need to consider only the two remaining cases.  

We first handle the case $q=p=2$ and state the following lemma.  

\begin{lemma}\label{lem-2DPDOGold} 
Let $q=p=2$. 
Let $s^{\infty}$ be the sequence of (\ref{eqn-sequence}), where $f(x)=D_3(x,a)=x^3-3ax=x^3+ax$.  
Then the minimal polynomial $\m_s(x)$ of  $s^{\infty}$ is given by 
\begin{eqnarray*}
\m_s(x)=\left\{ 
\begin{array}{ll} 
(x-1)^{\delta(1)} m_{\alpha^{-3}}(x)  & \mbox{if } a =0, \\
(x-1)^{\delta(1+a)} m_{\alpha^{-1}}(x) m_{\alpha^{-3}}(x) & \mbox{if } a \ne 0 
\end{array} 
\right. 
\end{eqnarray*} 
where $m_{\alpha^{-j}}(x)$ and the function $\delta(x)$ were defined in Section \ref{sec-notations}, 
and the linear span $\ls_s$ of $s^{\infty}$ is given by 
\begin{eqnarray*}
\ls_s=\left\{ 
\begin{array}{ll} 
\delta(1) + m   & \mbox{if } a =0, \\
\delta(1+a) + 2m & \mbox{if } a \ne 0.  
\end{array} 
\right. 
\end{eqnarray*} 
\end{lemma}

The following theorem gives information on the code $\C_{s}$.    

\begin{theorem}\label{thm-2DO2Gold} 
Let $q=p=2$ and let $m \geq 4$. 
Then the binary code $\C_{s}$ defined by the sequence of Lemma \ref{lem-2DPDOGold} has parameters 
$[n, n-\ls_s, d]$ and generator polynomial $\m_s(x)$, where $\m_s(x)$ and $\ls_s$ are 
given in Lemma \ref{lem-2DPDOGold}, and 
\begin{eqnarray*}
d=\left\{ 
\begin{array}{ll} 
2   & \mbox{if } a =0 \mbox{ and }  \delta(1)=0, \\
4   & \mbox{if } a =0 \mbox{ and }  \delta(1)=1, \\
5   & \mbox{if } a \ne 0 \mbox{ and }  \delta(1+a)=0, \\
6   & \mbox{if } a \ne 0 \mbox{ and }  \delta(1+a)=1. 
\end{array} 
\right. 
\end{eqnarray*} 
\end{theorem} 

\begin{remark} 
When $a=0$ and $\delta(1)=1$, the code is equivalent to the even-weight subcode of 
the Hamming code. We are mainly interested in the case that $a \ne 0$. When $a=1$, the code 
$\C_s$ is a double-error correcting binary BCH code or its even-like  
subcode. Theorem \ref{thm-2DO2Gold} shows that well-known classes of cyclic codes can be 
constructed with Dickson polynomials of order 3. The code is either optimal or almost optimal. 
\end{remark}

Now we consider the case $q=p^t$, where $p \ge 5$ or $p=2$ and $t\ge 2$. 

\begin{lemma}\label{lem-3tDPDOGold} 
Let $q=p^t$, where $p \ge 5$ or $p=2$ and $t\ge 2$. 
Let $s^{\infty}$ be the sequence of (\ref{eqn-sequence}), where $f(x)=D_3(x,a)=x^3-3ax$.  
Then the minimal polynomial $\m_s(x)$ of  $s^{\infty}$ is given by 
\begin{eqnarray*}
\m_s(x)=\left\{ 
\begin{array}{ll} 
(x-1)^{\delta(-2)} m_{\alpha^{-3}}(x)  m_{\alpha^{-2}}(x) & \mbox{if } a =1, \\
(x-1)^{\delta(1-3a)} m_{\alpha^{-3}}(x) m_{\alpha^{-2}}(x)  m_{\alpha^{-1}}(x) & \mbox{if } a \ne 1 
\end{array} 
\right. 
\end{eqnarray*} 
where $m_{\alpha^{-j}}(x)$ and the function $\delta(x)$ were defined in Section \ref{sec-notations}, 
and the linear span $\ls_s$ of $s^{\infty}$ is given by 
\begin{eqnarray*}
\ls_s=\left\{ 
\begin{array}{ll} 
\delta(-2) + 2m   & \mbox{if } a =1, \\
\delta(1+a) + 3m & \mbox{if } a \ne 1.  
\end{array} 
\right. 
\end{eqnarray*} 
\end{lemma}

The following theorem provides information on the code $\C_{s}$.    

\begin{theorem}\label{thm-3tDO2Gold} 
Let $q=p^t$, where $p \ge 5$ or $p=2$ and $t\ge 2$. 
Then the code $\C_{s}$ defined by the sequence of Lemma \ref{lem-3tDPDOGold} has parameters 
$[n, n-\ls_s, d]$ and generator polynomial $\m_s(x)$, where $\m_s(x)$ and $\ls_s$ are 
given in Lemma \ref{lem-3tDPDOGold}, and 
\begin{eqnarray*}
\left\{ 
\begin{array}{ll} 
d \ge 3   & \mbox{if } a =1, \\
d \ge 4   & \mbox{if } a \ne 1 \mbox{ and }  \delta(1-3a)=0, \\
d \ge 5   & \mbox{if } a \ne 1 \mbox{ and }  \delta(1-3a)=1,  \\  
d \ge 5   & \mbox{if } a \ne 1 \mbox{ and }  \delta(1-3a)=0 \mbox{ and } q=4, \\
d \ge 6   & \mbox{if } a \ne 1 \mbox{ and }  \delta(1-3a)=1 \mbox{ and } q=4.  
\end{array} 
\right. 
\end{eqnarray*} 
\end{theorem} 

\begin{remark} 
The code $\C_{s}$ of Theorem \ref{thm-3tDO2Gold} is either a BCH code or the even-like subcode 
of a BCH code. 
One can similarly show that the code is either optimal 
or almost optimal. 

When $q=4$, $a \neq 1$, $\delta(1-3a)=1$, and $m \geq 3$, the Sphere Packing Bound shows that 
$d=6$. But the minimum distance is still open in other cases.  
\end{remark} 

\begin{Open} 
Determine the minimum distance $d$ for the code $\C_{s}$ of Theorem \ref{thm-3tDO2Gold}. 
\end{Open}

\subsubsection{Cyclic codes from $D_4(x,a)=x^4-4ax^2+2a^2$}\label{sec-4DPDOGold} 

In this section we deal with the code $\C_s$ defined by the Dickson polynomial 
$D_4(x,a)=x^4-4ax^2+2a^2$. We have to distinguish among the three cases: $p=2$, 
$p=3$ and $p \ge 5$. The case $p=2$ was covered in Section \ref{sec-tracex}. 
So we need to consider only the two remaining cases.  

We first take care of the case $q=p=3$ and have the following lemma.  

\begin{lemma}\label{lem-34DPDOGold} 
Let $q=p=3$ and $m \ge 3$. 
Let $s^{\infty}$ be the sequence of (\ref{eqn-sequence}), where $f(x)=D_4(x,a)=x^4-4ax^2+2a^2$.  
Then the minimal polynomial $\m_s(x)$ of  $s^{\infty}$ is given by 
\begin{eqnarray*}
\m_s(x)=
 \left\{ 
\begin{array}{l} 
(x-1)^{\delta(1)} m_{\alpha^{-4}}(x)  m_{\alpha^{-1}}(x)  \mbox{ if } a =0, \\
(x-1)^{\delta(1)} m_{\alpha^{-4}}(x)  m_{\alpha^{-2}}(x)  \mbox{ if } a =1, \\
(x-1)^{\delta(1-a-a^2)} m_{\alpha^{-4}}(x)  m_{\alpha^{-2}}(x) m_{\alpha^{-1}}(x)  \mbox{ otherwise,}  
\end{array} 
\right. 
\end{eqnarray*} 
where $m_{\alpha^{-j}}(x)$ and the function $\delta(x)$ were defined in Section \ref{sec-notations}, 
and the linear span $\ls_s$ of $s^{\infty}$ is given by 
\begin{eqnarray*}\label{eqn-34DPDOGold}
\ls_s=\left\{ 
\begin{array}{ll} 
\delta(1) + 2m   & \mbox{if } a =0, \\
\delta(1) + 2m   & \mbox{if } a =1, \\
\delta(1-a-a^2) + 3m & \mbox{otherwise.}   
\end{array} 
\right. 
\end{eqnarray*} 
\end{lemma}

The following theorem gives information on the code $\C_{s}$.    

\begin{theorem}\label{thm-34DO2Gold} 
Let $q=p=3$ and $m \ge 3$. 
Then the code $\C_{s}$ defined by the sequence of Lemma \ref{lem-34DPDOGold} has parameters 
$[n, n-\ls_s, d]$ and generator polynomial $\m_s(x)$, where $\m_s(x)$ and $\ls_s$ are 
given in Lemma \ref{lem-34DPDOGold}, and 
\begin{eqnarray*}
\left\{ 
\begin{array}{ll} 
d = 2   & \mbox{if } a =1,  \\
d = 3   & \mbox{if } a =0  \mbox{ $m \equiv 0 \pmod{6}$,}   \\
d \ge 4   & \mbox{if } a =0  \mbox{ $m \not\equiv 0 \pmod{6}$,}   \\
d \ge 5   & \mbox{if } a^2 \ne a \mbox{ and }  \delta(1-a-a^2)=0, \\
d = 6   & \mbox{if } a^2 \ne a \mbox{ and }  \delta(1-a-a^2) =1. 
\end{array} 
\right. 
\end{eqnarray*} 
\end{theorem}

\begin{remark} 
When $a=1$, the code of Theorem \ref{thm-34DO2Gold} is neither optimal nor almost optimal.  
The code is either optimal or almost optimal in all other cases. 
\end{remark} 

\begin{Open} 
Determine the minimum distance $d$ for the code $\C_{s}$ of Theorem \ref{thm-34DO2Gold}. 
\end{Open}

Now we consider the case $q=p^t$, where $p \ge 5$ or $p=3$ and $t \ge 2$. 

\begin{lemma}\label{lem-4tDPDOGold} 
Let $m \ge 2$ and $q=p^t$, where $p \ge 5$ or $p=3$ and $t\ge 2$. 
Let $s^{\infty}$ be the sequence of (\ref{eqn-sequence}), where $f(x)=D_4(x,a)=x^4-4ax^2+2a^2$.  
Then the minimal polynomial $\m_s(x)$ of  $s^{\infty}$ is given by 
\begin{eqnarray*}
\m_s(x)=\left\{ 
\begin{array}{l} 
(x-1)^{\delta(1)} m_{\alpha^{-4}}(x)  m_{\alpha^{-3}}(x)  m_{\alpha^{-1}}(x)  \mbox{ if } a =\frac{3}{2}, \\
(x-1)^{\delta(1)} m_{\alpha^{-4}}(x)  m_{\alpha^{-3}}(x)  m_{\alpha^{-2}}(x)  \mbox{ if } a =\frac{1}{2}, \\
(x-1)^{\delta(1-4a+2a^2)} \prod_{i=1}^4 m_{\alpha^{-i}}(x)  \mbox{ if } a \not\in \{\frac{3}{2}, \frac{1}{2}\},   
\end{array} 
\right. 
\end{eqnarray*} 
where $m_{\alpha^{-j}}(x)$ and the function $\delta(x)$ were defined in Section \ref{sec-notations}, 
and the linear span $\ls_s$ of $s^{\infty}$ is given by 
\begin{eqnarray*}
\ls_s=\left\{ 
\begin{array}{ll} 
\delta(1) + 3m  & \mbox{if } a \in \{\frac{3}{2}, \frac{1}{2}\}, \\
\delta(1-4a+2a^2) + 4m  & \mbox{otherwise.}  
\end{array} 
\right. 
\end{eqnarray*} 
\end{lemma}

The following theorem delivers to us information on the code $\C_{s}$.    

\begin{theorem}\label{thm-4tDO2Gold} 
Let $m \ge 2$ and $q=p^t$, where $p \ge 5$ or $p=3$ and $t\ge 2$. 
Then the code $\C_{s}$ defined by the sequence of Lemma \ref{lem-4tDPDOGold} has parameters 
$[n, n-\ls_s, d]$ and generator polynomial $\m_s(x)$, where $\m_s(x)$ and $\ls_s$ are 
given in Lemma \ref{lem-4tDPDOGold}, and 
\begin{eqnarray*}
\left\{ 
\begin{array}{ll} 
d \ge 3   & \mbox{if } a =\frac{3}{2}, \\
d \ge 4   & \mbox{if } a =\frac{1}{2}, \\
d \ge 5   & \mbox{if } a \not\in \{\frac{3}{2}, \frac{1}{2}\}  \mbox{ and }  \delta(1-4a+a^2)=0, \\
d = 6   & \mbox{if } a \not\in \{\frac{3}{2}, \frac{1}{2}\}  \mbox{ and }  \delta(1-4a+a^2)=1. 
\end{array} 
\right. 
\end{eqnarray*} 
\end{theorem} 

\begin{remark} 
Except the cases that $a \in \{\frac{3}{2}, \frac{1}{2}\}$, the code $\C_{s}$ of Theorem 
\ref{thm-4tDO2Gold} is either optimal or almost optimal. 
\end{remark} 

\begin{Open} 
Determine the minimum distance $d$ for the code $\C_{s}$ of Theorem \ref{thm-4tDO2Gold}. 
\end{Open} 

\subsubsection{Cyclic codes from $D_5(x,a)=x^5-5ax^3+5a^2x$}\label{sec-5DOGold} 

In this section we deal with the code $\C_s$ defined by the Dickson polynomial 
$D_5(x,a)=x^5-5ax^3+5a^2x$. We have to distinguish among the three cases: $p=2$, 
$p=3$ and $p \ge 7$. The case $p=5$ was covered in Section \ref{sec-tracex}. 
So we need to consider only the remaining cases.  

We first consider the case $q=p=2$ and have the following lemma.  

\begin{lemma}\label{lem-52DPDOGold} 
Let $q=p=2$ and $m \ge 5$. 
Let $s^{\infty}$ be the sequence of (\ref{eqn-sequence}), where $f(x)=D_5(x,a)=x^5-5ax^3+5a^2x$.  
Then the minimal polynomial $\m_s(x)$ of  $s^{\infty}$ is given by 
\begin{eqnarray*}
\m_s(x)=  
 \left\{ 
\begin{array}{l} 
(x-1)^{\delta(1)} m_{\alpha^{-5}}(x)  \mbox{ if } a =0, \\
(x-1)^{\delta(1)} m_{\alpha^{-5}}(x)  m_{\alpha^{-3}}(x)  \mbox{ if } 1+a+a^3=0, \\
(x-1)^{\delta(1)} \prod_{i=0}^2 m_{\alpha^{-(2i+1)}}(x)   \mbox{ if } a+a^2+a^4 \ne 0 
\end{array} 
\right. 
\end{eqnarray*} 
where $m_{\alpha^{-j}}(x)$ and the function $\delta(x)$ were defined in Section \ref{sec-notations}, 
and the linear span $\ls_s$ of $s^{\infty}$ is given by 
\begin{eqnarray*}
\ls_s=\left\{ 
\begin{array}{ll} 
\delta(1) +m    &   \mbox{ if } a =0, \\
\delta(1) +2m  &   \mbox{ if } 1+a+a^3=0, \\
\delta(1) +3m  &   \mbox{ if } a+a^2+a^4 \ne 0.  
\end{array} 
\right. 
\end{eqnarray*} 
\end{lemma}

The following theorem describes parameters of the code $\C_{s}$.    

\begin{theorem}\label{thm-52DO2Gold} 
Let $q=p=2$ and $m \ge 5$. 
Then the code $\C_{s}$ defined by the sequence of Lemma \ref{lem-52DPDOGold} has parameters 
$[n, n-\ls_s, d]$ and generator polynomial $\m_s(x)$, where $\m_s(x)$ and $\ls_s$ are 
given in Lemma \ref{lem-52DPDOGold}, and 
\begin{eqnarray*}
\left\{ 
\begin{array}{ll} 
d = 2      & \mbox{if } a =0 \mbox{ and }  \delta(1)=0 \mbox{ and } \gcd(5, n)=5, \\
d = 3      & \mbox{if } a =0 \mbox{ and }  \delta(1)=0 \mbox{ and } \gcd(5, n)=1, \\
d = 4      & \mbox{if } a =0 \mbox{ and }  \delta(1)=1, \\
d \ge 3   & \mbox{if } 1+a+a^3=0 \mbox{ and }  \delta(1)=0, \\
d \ge 4   & \mbox{if } 1+a+a^3=0 \mbox{ and }  \delta(1)=1, \\
d \ge 7   & \mbox{if } a+a^2+a^4 \ne 0 \mbox{ and }  \delta(1)=0, \\
d = 8   & \mbox{if } a+a^2+a^4 \ne 0 \mbox{ and }  \delta(1)=1. 
\end{array} 
\right. 
\end{eqnarray*} 
\end{theorem}

\begin{remark} 
The code of Theorem \ref{thm-52DO2Gold} is either optimal or almost optimal. 
The code is not a BCH code when $1+a+a^3=0$, and a BCH code in the remaining cases. 
\end{remark} 

\begin{Open} 
Determine the minimum distance $d$ for the code $\C_{s}$ of Theorem \ref{thm-52DO2Gold} 
for the three open cases. 
\end{Open}

We now consider the case $(p, q)=(2, 4)$ and have the following lemma.  

\begin{lemma}\label{lem-452DPDOGold} 
Let $(p, q)=(2, 4)$ and $m \ge 3$. 
Let $s^{\infty}$ be the sequence of (\ref{eqn-sequence}), where $f(x)=D_5(x,a)=x^5-5ax^3+5a^2x$.  
Then the minimal polynomial $\m_s(x)$ of  $s^{\infty}$ is given by 
\begin{eqnarray*}
\m_s(x)=   
 \left\{ 
\begin{array}{ll} 
(x-1)^{\delta(1)} m_{\alpha^{-5}}(x)  & \mbox{ if } a =0, \\
(x-1)^{\delta(1)} m_{\alpha^{-5}}(x) m_{\alpha^{-3}}(x) m_{\alpha^{-2}}(x) & \mbox{ if } a=1, \\
(x-1)^{\delta(1+a+a^2)}  m_{\alpha^{-5}}(x) m_{\alpha^{-3}}(x) m_{\alpha^{-2}}(x) m_{\alpha^{-1}}(x)  
   & \mbox{ if } a+a^2 \ne 0 
\end{array} 
\right. 
\end{eqnarray*} 
where $m_{\alpha^{-j}}(x)$ and the function $\delta(x)$ were defined in Section \ref{sec-notations}, 
and the linear span $\ls_s$ of $s^{\infty}$ is given by 
\begin{eqnarray*}
\ls_s=\left\{ 
\begin{array}{ll} 
\delta(1) +m  &   \mbox{ if } a=0, \\
\delta(1) +3m  &   \mbox{ if } a=1, \\
\delta(1) +4m  &   \mbox{ if } a+a^2 \ne 0.  
\end{array} 
\right. 
\end{eqnarray*} 
\end{lemma}

The following theorem supplies information on the code $\C_{s}$.    

\begin{theorem}\label{thm-452DO2Gold} 
Let $(p,q)=(2, 4)$ and $m \ge 3$. 
Then the code $\C_{s}$ defined by the sequence of Lemma \ref{lem-452DPDOGold} has parameters 
$[n, n-\ls_s, d]$ and generator polynomial $\m_s(x)$, where $\m_s(x)$ and $\ls_s$ are 
given in Lemma \ref{lem-452DPDOGold}, and 
\begin{eqnarray*}
\left\{ 
\begin{array}{ll} 
d = 2      & \mbox{if } a =0 \mbox{ and }  \delta(1)=0 \mbox{ and } \gcd(5, n)=5, \\
d = 3   & \mbox{if } a =0 \mbox{ and }  \gcd(5, n)=1, \\
d \ge 3   & \mbox{if } a=1, \\ 
d \ge 6   & \mbox{if } a+a^2 \ne 0 \mbox{ and }  \delta(1)=0, \\
d \ge 7   & \mbox{if } a+a^2 \ne 0 \mbox{ and }  \delta(1)=1.  
\end{array} 
\right. 
\end{eqnarray*} 
\end{theorem}

Examples of the code of Theorem \ref{thm-452DO2Gold} are documented in arXiv:1206.4370, 
and many of them are optimal. 

\begin{Open} 
Determine the minimum distance $d$ of the code $\C_s$ in Theorem \ref{thm-452DO2Gold}. 
\end{Open}

We now consider the case $(p, q)=(2, 2^t)$, where $t \ge 3$, and state the following lemma.  

\begin{lemma}\label{lem-t52DPDOGold} 
Let $(p, q)=(2, 2^t)$ and $m \ge 3$, where $t \ge 3$. 
Let $s^{\infty}$ be the sequence of (\ref{eqn-sequence}), where $f(x)=D_5(x,a)=x^5-5ax^3+5a^2x$.  
Then the minimal polynomial $\m_s(x)$ of  $s^{\infty}$ is given by 
\begin{eqnarray*}
\m_s(x)=    
 \left\{ 
\begin{array}{ll} 
(x-1)^{\delta(1)} m_{\alpha^{-5}}(x)  m_{\alpha^{-4}}(x)  m_{\alpha^{-1}}(x) & \mbox{ if } a =0, \\
\prod_{i=2}^5 m_{\alpha^{-i}}(x) &  \mbox{ if } 1+a+a^2=0, \\
(x-1)^{\delta(1+a+a^2)} \prod_{i=1}^5 m_{\alpha^{-i}}(x)  & \mbox{ if } a+a^2+a^3 \ne 0, 
\end{array} 
\right. 
\end{eqnarray*} 
where $m_{\alpha^{-j}}(x)$ and the function $\delta(x)$ were defined in Section \ref{sec-notations}, 
and the linear span $\ls_s$ of $s^{\infty}$ is given by 
\begin{eqnarray*}
\ls_s=\left\{ 
\begin{array}{ll} 
\delta(1) +3m  &   \mbox{ if } a=0, \\
\delta(1) +4m  &   \mbox{ if } 1+a+a^2=0, \\
\delta(1) +5m  &   \mbox{ if } a+a^2+a^3 \ne 0.  
\end{array} 
\right. 
\end{eqnarray*} 
\end{lemma}

The following theorem provides information on the code $\C_{s}$.    

\begin{theorem}\label{thm-t52DO2Gold} 
Let $(p, q)=(2, 2^t)$, where $t \ge 3$. 
Then the code $\C_{s}$ defined by the sequence of Lemma \ref{lem-t52DPDOGold} has parameters 
$[n, n-\ls_s, d]$ and generator polynomial $\m_s(x)$, where $\m_s(x)$ and $\ls_s$ are 
given in Lemma \ref{lem-t52DPDOGold}, and 
\begin{eqnarray*}
\left\{ 
\begin{array}{ll} 
d \ge 3   & \mbox{if } a =0 \mbox{ and }  \delta(1)=0,  \\
d \ge 4      & \mbox{if } a =0 \mbox{ and }  \delta(1)=1, \\
d \ge 5   & \mbox{if } 1+a+a^2=0, \\
d \ge 6   & \mbox{if } a+a^2+a^3 \ne 0 \mbox{ and }  \delta(1)=0, \\
d \ge 7   & \mbox{if } a+a^2+a^3 \ne 0 \mbox{ and }  \delta(1)=1. 
\end{array} 
\right. 
\end{eqnarray*} 
\end{theorem}

\begin{Open} 
Determine the minimum distance $d$ of the code $\C_s$ in Theorem \ref{thm-t52DO2Gold}. 
\end{Open} 

Examples of the code of Theorem \ref{thm-t52DO2Gold} can be found in arXiv:1206.4370, 
and many of them are optimal. The code of Theorem \ref{thm-t52DO2Gold} is 
not a BCH code when $a=0$, and a BCH code otherwise.    

We now consider the case $q=p=3$ and  state the following lemma and theorem.  

\begin{lemma}\label{lem-352DPDOGold} 
Let $q=p=3$ and $m \ge 3$. 
Let $s^{\infty}$ be the sequence of (\ref{eqn-sequence}), where $f(x)=D_5(x,a)=x^5-5ax^3+5a^2x$.  
Then the minimal polynomial $\m_s(x)$ of  $s^{\infty}$ is given by 
\begin{eqnarray*}
\m_s(x)=  
 \left\{ 
\begin{array}{ll} 
(x-1)^{\delta(1+a+2a^2)} m_{\alpha^{-5}}(x)  m_{\alpha^{-4}}(x) m_{\alpha^{-2}}(x) 
   & \mbox{ if } a - a^6 =0, \\
(x-1)^{\delta(1+a+2a^2)} \prod_{i=2}^5 m_{\alpha^{-i}}(x)  
   & \mbox{ if } a -a^6 \ne 0, 
\end{array} 
\right. 
\end{eqnarray*} 
where $m_{\alpha^{-j}}(x)$ and the function $\delta(x)$ were defined in Section \ref{sec-notations}, 
and the linear span $\ls_s$ of $s^{\infty}$ is given by 
\begin{eqnarray*}
\ls_s=\left\{ 
\begin{array}{ll} 
\delta(1+a+2a^2) + 3m    &   \mbox{ if } a - a^6 =0, \\
\delta(1+a+2a^2) + 4m    &   \mbox{ if } a - a^6 \ne 0. 
\end{array} 
\right. 
\end{eqnarray*} 
\end{lemma}

The following theorem gives information on the code $\C_{s}$.    

\begin{theorem}\label{thm-352DO2Gold} 
Let $q=p=3$ and $m \ge 3$. 
Then the code $\C_{s}$ defined by the sequence of Lemma \ref{lem-352DPDOGold} has parameters 
$[n, n-\ls_s, d]$ and generator polynomial $\m_s(x)$, where $\m_s(x)$ and $\ls_s$ are 
given in Lemma \ref{lem-352DPDOGold}, and 
\begin{eqnarray*}
\left\{ 
\begin{array}{ll} 
d \ge 4   & \mbox{if } a-a^6=0,  \\
d \ge 7   & \mbox{if } a-a^6\ne 0 \mbox{ and }  \delta(1+a+2a^2)=0, \\
d \ge 8   & \mbox{if } a-a^6\ne 0 \mbox{ and }  \delta(1+a+2a^2)=1. 
\end{array} 
\right. 
\end{eqnarray*} 
\end{theorem} 

\begin{Open} 
Determine the minimum distance $d$ of the code $\C_s$ in Theorem \ref{thm-352DO2Gold} (our experimental data 
indicates that the lower bounds are the specific values of $d$). 
\end{Open}

Examples of the code of Theorem \ref{thm-352DO2Gold} are described in arXiv:1206.4370, 
and some of them are optimal.

We now consider the case $(p, q)=(3, 3^t)$, where $t \ge 3$, and state the following lemma and theorem.  

\begin{lemma}\label{lem-t53DPDOGold} 
Let $(p, q)=(3, 3^t)$ and $m \ge 2$, where $t \ge 2$. 
Let $s^{\infty}$ be the sequence of (\ref{eqn-sequence}), where $f(x)=D_5(x,a)=x^5-5ax^3+5a^2x$.  
Then the minimal polynomial $\m_s(x)$ of  $s^{\infty}$ is given by 
\begin{eqnarray*}
\m_s(x)=    
 \left\{ 
\begin{array}{ll} 
(x-1)^{\delta(1)} m_{\alpha^{-5}}(x)  m_{\alpha^{-4}}(x)  m_{\alpha^{-2}}(x) m_{\alpha^{-1}}(x) 
  & \mbox{ if } 1+a =0, \\
(x-1)^{\delta(a-1)} m_{\alpha^{-5}}(x) m_{\alpha^{-4}}(x) m_{\alpha^{-3}}(x) m_{\alpha^{-2}}(x)  
  & \mbox{ if } 1+a^2=0, \\
(x-1)^{\delta(1+a+2a^2)} \prod_{i=1}^5 m_{\alpha^{-i}}(x)  
  &  \mbox{ if } (a+1)(a^2+1) \ne 0, 
\end{array} 
\right. 
\end{eqnarray*} 
where $m_{\alpha^{-j}}(x)$ and the function $\delta(x)$ were defined in Section \ref{sec-notations}, 
and the linear span $\ls_s$ of $s^{\infty}$ is given by 
\begin{eqnarray*}
\ls_s=\left\{ 
\begin{array}{ll} 
\delta(1) +4m  &   \mbox{ if } a+1=0, \\
\delta(a-1) +4m  &   \mbox{ if } a^2+1=0, \\
\delta(1+a+2a^2) +5m  &   \mbox{ if } (a+1)(a^2+1) \ne 0.  
\end{array} 
\right. 
\end{eqnarray*} 
\end{lemma}

The following theorem supplies information on the code $\C_{s}$.    

\begin{theorem}\label{thm-t53DO2Gold} 
Let $(p, q)=(3, 3^t)$ and $m \ge 2$, where $t \ge 2$. 
Then the code $\C_{s}$ defined by the sequence of Lemma \ref{lem-t52DPDOGold} has parameters 
$[n, n-\ls_s, d]$ and generator polynomial $\m_s(x)$, where $\m_s(x)$ and $\ls_s$ are 
given in Lemma \ref{lem-t53DPDOGold}, and 
\begin{eqnarray*}
\left\{ 
\begin{array}{ll} 
d \ge 3      & \mbox{if } a =-1 \mbox{ and }  \delta(1)=0,  \\
d \ge 4      & \mbox{if } a =-1 \mbox{ and }  \delta(1)=1,  \\
d \ge 5      & \mbox{if } a^2 =-1 \mbox{ and }  \delta(a-1)=0, \\
d \ge 6      & \mbox{if } a^2 =-1 \mbox{ and }  \delta(a-1)=1, \\
d \ge 6      & \mbox{if } (a+1)(a^2+1) \ne 0 \mbox{ and }  \delta(1+a+2a^2)=0, \\
d \ge 7      & \mbox{if } (a+1)(a^2+1)  \ne 0 \mbox{ and }  \delta(1+a+2a^2)=1. 
\end{array} 
\right. 
\end{eqnarray*} 
\end{theorem}

\begin{Open} 
Determine the minimum distance $d$ of the code $\C_s$ in Theorem \ref{thm-t53DO2Gold}. 
\end{Open} 

Examples of the code of Theorem \ref{thm-t53DO2Gold} are available in  arXiv:1206.4370, 
and some of them are optimal. The code is a BCH code, except in the case that $a=-1$.

We finally consider the case $p \ge 7$, and present the following lemma and theorem.  

\begin{lemma}\label{lem-tt752DPDOGold} 
Let $p \ge 7$ and $m \geq 2$. 
Let $s^{\infty}$ be the sequence of (\ref{eqn-sequence}), where $f(x)=D_5(x,a)=x^5-5ax^3+5a^2x$.  
Then the minimal polynomial $\m_s(x)$ of  $s^{\infty}$ is given by 
\begin{eqnarray*}
\m_s(x)=    
 \left\{ 
\begin{array}{l} 
(x-1)^{\delta(1-5a+5a^2)} m_{\alpha^{-5}}(x)  m_{\alpha^{-4}}(x)  m_{\alpha^{-2}}(x) m_{\alpha^{-1}}(x) 
   \mbox{ if } a =2, \\
(x-1)^{\delta(1-5a+5a^2)} m_{\alpha^{-5}}(x) m_{\alpha^{-4}}(x) m_{\alpha^{-3}}(x) m_{\alpha^{-1}}(x) 
     \mbox{ if } a =\frac{2}{3}, \\
(x-1)^{\delta(1-5a+5a^2)} m_{\alpha^{-5}}(x) m_{\alpha^{-4}}(x) m_{\alpha^{-3}}(x) m_{\alpha^{-2}}(x)  
     \mbox{ if } a^2-3a+1=0, \\
(x-1)^{\delta(1-5a+5a^2)} \prod_{i=1}^5 m_{\alpha^{-i}}(x)   
  \mbox{ if } (a^2-3a+1)(a-2)(3a-2) \ne 0,  
\end{array} 
\right. 
\end{eqnarray*} 
where $m_{\alpha^{-j}}(x)$ and the function $\delta(x)$ were defined in Section \ref{sec-notations}, 
and the linear span $\ls_s$ of $s^{\infty}$ is given by 
\begin{eqnarray*}
\ls_s=\left\{ 
\begin{array}{l} 
\delta(1-5a+5a^2) +4m,     \mbox{ if } (a^2-3a+1)(a-2)(3a-2) =0, \\
\delta(1-5a+5a^2) +5m,  \mbox{ otherwise. }   
\end{array} 
\right. 
\end{eqnarray*} 
\end{lemma}

The following theorem provides information on the code $\C_{s}$.    

\begin{theorem}\label{thm-tt752DO2Gold} 
Let $p\ge 7$ and $m \geq 2$. 
Then the code $\C_{s}$ defined by the sequence of Lemma \ref{lem-tt752DPDOGold} has parameters 
$[n, n-\ls_s, d]$ and generator polynomial $\m_s(x)$, where $\m_s(x)$ and $\ls_s$ are 
given in Lemma \ref{lem-tt752DPDOGold}, and 
\begin{eqnarray*}
\left\{ 
\begin{array}{ll} 
d \ge 3   & \mbox{if } a =2 \mbox{ and }  \delta(1-5a+5a^2)=0,  \\
d \ge 4   & \mbox{if } a =2 \mbox{ and }  \delta(1-5a+5a^2)=1, \\
d \ge 4   & \mbox{if } a =\frac{2}{3} \mbox{ and }  \delta(1-5a+5a^2)=0,  \\
d \ge 5   & \mbox{if } a =\frac{2}{3} \mbox{ and }  \delta(1-5a+5a^2)=1, \\
d \ge 5   & \mbox{if } 1-3a+a^2=0 \mbox{ and }  \delta(1-5a+5a^2)=0, \\
d \ge 6   & \mbox{if } 1-3a+a^2=0 \mbox{ and }  \delta(1-5a+5a^2)=1, \\
d \ge 6   & \mbox{if } (a^2-3a+1)(a-2)(3a-2)  \ne 0 \mbox{ and }  
               \delta(1-5a+5a^2)=0, \\
d \ge 7   & \mbox{if } (a^2-3a+1)(a-2)(3a-2)  \ne 0 \mbox{ and }  
               \delta(1-5a+5a^2)=1. 
\end{array} 
\right. 
\end{eqnarray*} 
\end{theorem} 

\begin{Open} 
Determine the minimum distance $d$ of the code $\C_s$ in Theorem \ref{thm-tt752DO2Gold}. 
\end{Open}

Examples of the code of Theorem \ref{thm-tt752DO2Gold} can be found in  arXiv:1206.4370, 
and some of them are optimal. The code is a BCH code, except in the cases $a \in \{2, 2/3\}$.

\subsubsection{Cyclic codes from other $D_i(x,a)$ for $i \ge 6$} 

Parameters of cyclic codes from $D_i(x,a)$ for small $i$ could be established in a very similar 
way. However, more cases are involved and the situation is getting more complicated 
when $i$ gets bigger. Examples of the code $\C_s$ from $D_7(x,a)$ and $D_{11}(x,a)$ 
can be found in arXiv:1206.4370.

\subsubsection{Cyclic codes from Dickson polynomials of the second kind}\label{sec-2ndDPcode} 

Results on cyclic codes from Dickson polynomials of the second kind can be developed in 
a similar way. 
Experimental data indicates that the codes from the Dickson polynomials of the first kind are 
in general better than those from the Dickson polynomials of the second kind, though some 
cyclic codes from Dickson polynomials of the second kind could also be optimal or almost 
optimal.   

\subsubsection{Comments on the cyclic codes from Dickson polynomials}

It is really amazing that in most cases the cyclic codes derived from the Dickson 
polynomials of small degrees within the framework of this paper are optimal or almost optimal (see arXiv:1206.4370 for examples of optimal codes).  
  
We had to treat Dickson polynomials of small degrees case by case over finite fields with different 
characteristics as we do not see a way of treating them in a single strike. The generator polynomial 
and the dimension of the codes depend heavily on the degree of the Dickson polynomials and the 
characteristic of the base field. 

\section{Cyclic codes from the two-prime sequences}\label{sec-2pqseq} 

The objective of this section is to document a family of codes from the two-prime sequences. 
All the materials of this section come from \cite{Ding120}. 

\subsection{The two-prime sequences} 

Throughout this section,  let $n_1$ and ${n_2}$ be two distinct odd primes, define $n=n_1n_2$ and 
\begin{eqnarray*} 
{N_1}=\{n_1, {2n_1}, \cdots, ({n_2}-1)n_1\}, \ {N_2}=\{{n_2},2n_2,\cdots, (n_1-1){n_2}\}.  
\end{eqnarray*} 
The two-prime sequence, denoted by $\lambda^{\infty}$ throughout this section exclusively,  
is defined by 
\begin{eqnarray}\label{eqn-gotor} 
\lambda_i=\left\{ \begin{array}{ll} 
             0, & \mbox { if } i \bmod n \in \{0\} \cup {N_2}, \\ 
             1, &  \mbox { if } i \bmod n \in {N_1}, \\ 
             \left(1-\left(\frac{i}{n_1}\right)\left(\frac{i}{{n_2}}\right)\right)/2
                & \mbox{otherwise,} 
            \end{array} \right. 
\end{eqnarray}  
where $\left(\frac{a}{n_1}\right)$ denotes the Legendre symbol. The autocorrelation values  
of this binary sequence $\lambda^{\infty}$  were determined in \cite{Ding98ffa,German}. Traditionally, these two-prime sequences are 
defined as binary sequences and a generalization of the binary twin-prime sequences \cite{Whit62}. 
Here in this section, we treat them as sequences over any finite 
field $\gf(q)$, where $\gcd(q, n)=1$, and will use them to construct cyclic codes of length 
$n$ over $\gf(q)$. 

We now give a cyclotomic description of the sequence $\lambda^{\infty}$. 
An integer $a$ is called a primitive root of (or modulo) $n$ if the 
multiplicative order of $a$ modulo $n$, denoted by $\ord_n(a)$, is 
equal to $\phi(n)$, where $\phi(x)$ is the Euler function and 
$\gcd(a, n)=1$. 

Define $N=\gcd(n_1-1, n_2-1)$ and 
$e=(n_1-1)({n_2}-1)/{N}$. It is well-known that any prime $n_1$ has $\phi(n_1-1)$ 
primitive roots. The Chinese Remainder Theorem guarantees that there 
are common primitive roots of both $n_1$ and ${n_2}$. Let ${\pi}$ be a fixed 
common primitive root of both $n_1$ and ${n_2}$, and $ \varrho$ be an integer 
satisfying 
\begin{eqnarray*} 
\begin{array}{ll} 
\varrho \equiv {\pi} \pmod {n_1}, \ \ \varrho \equiv 1 \pmod {n_2}. 
\end{array} 
\end{eqnarray*} 
Whiteman proved that \cite{Whit62} 
\begin{eqnarray*} 
\Z_{n}^*=\{{\pi}^s \varrho^i: s=0,1, \cdots, e-1; \ i=0, 1, \cdots, {N}-1\}, 
\end{eqnarray*} 
where $\Z_{n}^*$ denotes the set of all invertible elements of the residue 
class ring $\Z_{n}$. The generalized cyclotomic classes $W_i$ of order ${N}$ 
with respect to $n_1$ and ${n_2}$ are defined by 
\begin{eqnarray*} 
W_i^{(N)}=\{{\pi}^s \varrho^i: s=0,1, \cdots, e-1\}, \ \ i=0, 1, \cdots, {N}-1.   
\end{eqnarray*} 
It was proved in \cite{Whit62} that   
\begin{eqnarray*} 
\Z_{n}^*=\cup_{i=0}^{{N}-1} W_i^{(N)}, \ \ W_i^{(N)} \cap W_j^{(N)} =\emptyset \mbox{ for } i \neq j. 
\end{eqnarray*} 
This generalized cyclotomy was 
introduced by Whiteman \cite{Whit62}. 
The motivation behind the investigation of the generalized cyclotomy with 
respect to two primes is the search for residue difference sets. The famous 
twin-prime difference sets are among such a class of difference sets. 

Define 
$$ 
D_0^{(2)}=\cup_{i=0}^{(N-2)/2} W_{2i}^{(N)} \mbox{ and }  D_1^{(2)}=\cup_{i=0}^{(N-2)/2} W_{2i+1}^{(N)}. 
$$
Clearly $D_0^{(2)}$ is a subgroup of $\Z_n^*$ and $D_1^{(2)}=\varrho D_0^{(2)}$. The sets $D_0^{(2)}$  and 
$D_1^{(2)}$ are called the extended generalized cyclotomic classes of order two and are identical to 
Whiteman's cyclotomic classes of order $N$ when and only when $N=2$.  In other words, the cyclotomy 
$\{D_0^{(2)},  D_1^{(2)}\}$ is different from Whiteman's cyclotomy when $N>2$.  

Let 
$$ 
C_0=\{0\}  \cup {N_2} \cup D_0^{(2)}, \ C_1={N_1} \cup D_1^{(2)}.  
$$
Then 
\begin{eqnarray*} 
C_0 \cup C_1 =\Z_{{n}}, \ \ C_0 \cap C_1 = \emptyset. 
\end{eqnarray*} 
It is not hard to verify that 
\begin{eqnarray*} 
\lambda_i=\left\{ \begin{array}{ll} 
             0, & \mbox{ if } i \bmod n \in C_0 \\ 
             1, & \mbox{ if } i \bmod n \in C_1.  
            \end{array} \right. 
\end{eqnarray*} 
This is the cyclotomic description of the two-prime sequence $\lambda^{\infty}$ defined in (\ref{eqn-gotor}). 

In certain special cases, the two-prime sequence has optimal autocorrelation \cite{Ding98}. As will be seen later, 
the code $\C_{\lambda}$ has also very good parameters in special cases. 

\subsection{The codes $\C_{\lambda}$ defined by the two-prime sequences $\lambda^{\infty}$} 

After the preparations above, we are ready to present the linear span and minimal polynomial 
of the sequence $\lambda^{\infty}$ defined in (\ref{eqn-gotor}). To this end, we need to discuss 
the factorization of $x^n-1$ over $\gf(q)$. 

Let $\theta$ be the same as before. Among the ${n}$th roots of unity 
$\theta^i$, where $0 \leq i \leq {n}-1$, the ${n_2}$ elements $\theta^i$, 
$i \in {N_1} \cup \{0\}$, are ${n_2}$th roots of unity, the ${n_1}$ elements $\theta^i$, 
$i \in {N_2} \cup \{0\}$, are ${n_1}$th roots of unity. Hence, 
\begin{eqnarray*} 
x^{n_1}-1=\prod_{i \in {N_2} \cup \{0\}} (x-\theta^i), \ \ 
x^{n_2}-1=\prod_{i \in {N_1} \cup \{0\}} (x-\theta^i).  
\end{eqnarray*} 

We define 
$$ 
d_i(x)=\prod_{i \in D_i^{(2)}} (x-\theta^i)  
$$
for $i \in \{0,1\}$. If $q \in D_0^{(2)}$, it is easily proved that $d_i(x) \in \gf(q)[x]$ 
for all $i$. 

Let $d(x)=d_0(x)d_1(x) \in \gf(q)[x]$.  We have then 
\begin{eqnarray*} 
x^{{n}}-1=\prod_{i=0}^{{n}-1} (x-\theta^i)=\frac{(x^{n_1}-1)(x^{n_2}-1)}{x-1} d(x).   
\end{eqnarray*} 

\begin{theorem}\label{thm-mmm}
Define for $i \in \{1,2\}$
\begin{eqnarray*} 
\Delta_i=\frac{n_i+(-1)^{i-1}}{2} \bmod{p}, \   
\Delta=\frac{(n_1+1)(n_2-1)}{2} \bmod{p}. 
\end{eqnarray*}
\begin{enumerate}
\item 
When $n \equiv 3 \pmod{4}$ and $\frac{n+1}{4} \bmod{p} \ne 0$ or $n \equiv 1 \pmod{4}$ and $\frac{n-1}{4} \bmod{p} \ne 0$,  
the minimal polynomial of the sequence 
$\lambda^{\infty}$ is given by 
\begin{eqnarray}\label{eqn-mini909} 
\m_\lambda(x)=  
\left\{ 
\begin{array}{l} 
x^n-1,  \mbox{ if }  \Delta_1 \ne 0, \ \Delta_2 \ne 0, \ \Delta \ne 0 \\
\frac{x^n-1}{x-1},  \mbox{ if } \Delta_1 \ne 0,  \ \Delta_2 \ne 0, \ \Delta = 0 \\ 
\frac{x^n-1}{x^{n_2}-1},  \mbox{ if } \Delta_1 = 0,  \ \Delta_2 \ne 0 \\ 
\frac{x^n-1}{x^{n_1}-1},  \mbox{ if } \Delta_1 \ne 0,  \ \Delta_2 = 0 \\ 
\frac{(x^n-1)(x-1)}{(x^{(n_1}-1) (x^{(n_2}-1)},  \mbox{ if } \Delta_1 = 0,  \ \Delta_2 = 0.  
\end{array} 
\right.
\end{eqnarray}
The linear span of the sequence $\lambda^{\infty}$ is equal to $\deg(\m_\lambda(x))$. 
In this case, the cyclic code $\calC_\lambda$ over 
          $\gf(q)$ defined by the sequence $\lambda^{\infty}$ has parameters $\left[n, k, d \right]$ and 
generator polynomial $\m_\lambda(x)$ of (\ref{eqn-mini909}), 
where the dimension $k=n-\deg(\m_\lambda(x))$.   

\item When $n \equiv 3 \pmod{4}$ and $\frac{n+1}{4} \bmod{p} = 0$ or $n \equiv 1 \pmod{4}$ and $\frac{n-1}{4} \bmod{p} = 0$,  
and $d_i(x) \in \gf(q)[x]$ 
for each $i$. In this case,  $\Lambda(\theta) \in \{0,1\}$ and the minimal polynomial of the sequence 
$\lambda^{\infty}$ is given by 
\begin{eqnarray}\label{eqn-mini910} 
\lefteqn{\m_\lambda(x)=} \nonumber \\
& 
\left\{ 
\begin{array}{l} 
\frac{x^n-1}{d_0(x)}, \mbox{ if } \Delta_1 \ne 0,  \ \Delta_2 \ne 0, \ \Delta \ne 0, \ \Lambda(\theta)=0 \\
\frac{x^n-1}{d_1(x)},  \mbox{ if } \Delta_1 \ne 0,  \ \Delta_2 \ne 0, \ \Delta \ne 0, \ \Lambda(\theta)=-1 \\
\frac{x^n-1}{(x-1)d_0(x)},  \mbox{ if } \Delta_1 \ne 0,  \ \Delta_2 \ne 0, \ \Delta = 0, \ \Lambda(\theta)=0 \\ 
\frac{x^n-1}{(x-1)d_1(x)},  \mbox{ if } \Delta_1 \ne 0,  \Delta_2 \ne 0,  \Delta = 0,  \Lambda(\theta)=-1 \\ 
\frac{x^n-1}{(x^{n_2}-1)d_0(x)},  \mbox{ if } \Delta_1 = 0,  \ \Delta_2 \ne 0, \ \Lambda(\theta)=0 \\ 
\frac{x^n-1}{(x^{n_2}-1)d_1(x)},  \mbox{ if } \Delta_1 = 0,  \ \Delta_2 \ne 0,  \ \Lambda(\theta)=-1 \\ 
\frac{x^n-1}{(x^{n_1}-1)d_0(x)},  \mbox{ if } \Delta_1 \ne 0,  \ \Delta_2 = 0, \ \Lambda(\theta)=0 \\ 
\frac{x^n-1}{(x^{n_1}-1)d_1(x)},  \mbox{ if } \Delta_1 \ne 0,  \ \Delta_2 = 0, \ \Lambda(\theta)=-1 \\ 
\frac{(x^n-1)(x-1)}{d_0(x)\prod_{i=1}^2(x^{n_i}-1)},  \mbox{ if } \Delta_1 = 0,  \ \Delta_2 = 0, \ \Lambda(\theta)=0 \\
\frac{(x^n-1)(x-1)}{d_1(x)\prod_{i=1}^2(x^{n_i}-1)},  \mbox{ if } \Delta_1 = 0,  \Delta_2 = 0,  \Lambda(\theta)=-1.  
\end{array} 
\right. 
\end{eqnarray}
The linear span of the sequence $\lambda^{\infty}$ is equal to $\deg(\m_\lambda(x))$.  
In this case, the cyclic code $\calC_\lambda$ over 
          $\gf(q)$ defined by the sequence $\lambda^{\infty}$ has parameters $\left[n, k, d \right]$ and 
generator polynomial $\m_\lambda(x)$ of (\ref{eqn-mini910}), 
where the dimension $k=n-\deg(\m_\lambda(x))$.   

\end{enumerate}
\end{theorem}

\subsubsection{The case $q=2$}

The following corollary follows directly from  Theorem \ref{thm-mmm}, 
and its conclusions on the linear span of the sequence $\lambda^\infty$ is an extension of the work 
in \cite{Ding98ffa}. 

\begin{corollary}\label{cor-lsq=2} 
Let $q=2$. We have the following conclusions: 
\begin{enumerate}
\item If ${n_1} \equiv 1 \pmod 8$ and ${n_2} \equiv 3 \pmod 8$ or ${n_1} \equiv 
           -3 \pmod 8$ and ${n_2} \equiv -1 \pmod 8$, we have  
           \begin{eqnarray*} 
            \ls_{\lambda}={n}-1, \ \ \m_\lambda(x)=\frac{x^{{n}}-1}{x-1}. 
           \end{eqnarray*} 
           In this case, the cyclic code $\calC_\lambda$ over 
          $\gf(q)$ defined by the sequence $\lambda^{\infty}$ has generator polynomial $\m_\lambda(x)$  
          above and parameters $\left[n, 1, n-1 \right]$.  

\item If ${n_1} \equiv -1 \pmod 8$ and ${n_2} \equiv 3 \pmod 8$ or ${n_1} \equiv 
           3 \pmod 8$ and ${n_2} \equiv -1 \pmod 8$, we have  
           \begin{eqnarray*} 
            \ls_{\lambda}=n-n_2, \ \ \m_\lambda(x)=\frac{x^{{n}}-1}{x^{n_2}-1}. 
           \end{eqnarray*} 
           In this case, the cyclic code $\calC_\lambda$ over 
          $\gf(q)$ defined by the sequence $\lambda^{\infty}$ has generator polynomial $\m_\lambda(x)$  
          above and parameters $\left[n, n_2, n_1 \right]$, where the minimum weight follows 
          from Theorem \ref{thm-2Pbound151}.   
                     
\item If ${n_1} \equiv -3 \pmod 8$ and ${n_2} \equiv 1 \pmod 8$ or ${n_1} \equiv 
           1 \pmod 8$ and ${n_2} \equiv -3 \pmod 8$, we have  
           \begin{eqnarray*} 
            \ls_{\lambda}=n-n_1, \ \ \m_\lambda(x)=\frac{x^{{n}}-1}{x^{n_1}-1}. 
           \end{eqnarray*}       
           In this case, the cyclic code $\calC_\lambda$ over 
          $\gf(q)$ defined by the sequence $\lambda^{\infty}$ has generator polynomial $\m_\lambda(x)$  
          above and parameters $\left[n, n_1, n_2 \right]$, where the minimum weight follows 
          from Theorem \ref{thm-2Pbound151}.               
                
\item If ${n_1} \equiv -1 \pmod 8$ and ${n_2} \equiv -3 \pmod 8$ or ${n_1} \equiv 
           3 \pmod 8$ and ${n_2} \equiv 1 \pmod 8$, we have  
           \begin{eqnarray*} 
            \ls_{\lambda}={n}-({n_1}+{n_2}-1), \ \ \m_\lambda(x)=\frac{(x^{{n}}-1)(x-1)}{(x^{n_2}-1)(x^{n_2}-1)}. 
           \end{eqnarray*}  
           In this case, the cyclic code $\calC_\lambda$ over 
          $\gf(q)$ defined by the sequence $\lambda^{\infty}$ has generator polynomial $\m_\lambda(x)$  
          above and parameters $\left[n, {n_1}+{n_2}-1, d \right]$, where $d$ has the lower bound 
          of Theorem \ref{thm-2Pbound152}.          
           
\item If ${n_1} \equiv 1 \pmod 8$ and ${n_2} \equiv -1 \pmod 8$ or ${n_1} \equiv 
           -3 \pmod 8$ and ${n_2} \equiv 3 \pmod 8$, we have  
           \begin{eqnarray*} 
& &            \ls_{\lambda}=n-\frac{(n_1-1)(n_2-1)+2}{2}, \\   
 & &           \m_\lambda(x)=\left\{ \begin{array}{ll} 
                                               \frac{x^{{n}}-1}{(x-1)d_0(x)}, & \mbox{ if } \Lambda(\theta)=0 \\
                                               \frac{x^{{n}}-1}{(x-1)d_1(x)}, & \mbox{ if } \Lambda(\theta)=1.  
                                               \end{array} 
                                    \right.                                                            
           \end{eqnarray*} 
           In this case, the cyclic code $\calC_\lambda$ over 
          $\gf(q)$ defined by the sequence $\lambda^{\infty}$ has generator polynomial $\m_\lambda(x)$  
          above and parameters $\left[n, k, d \right]$, where $k=\frac{(n_1-1)(n_2-1)+2}{2}$.             
           
\item If ${n_1} \equiv -1 \pmod 8$ and ${n_2} \equiv 1 \pmod 8$ or ${n_1} \equiv 
           3 \pmod 8$ and ${n_2} \equiv -3 \pmod 8$, we have  
           \begin{eqnarray*} 
             \ls_\lambda=n-\frac{(n_1+1)(n_2+1)-2}{2}   
           \end{eqnarray*} 
           and 
           \begin{eqnarray*} 
            \m_\lambda(x)=\left\{ \begin{array}{ll} 
            \frac{(x^n-1)(x-1)}{(x^{n_1}-1)(x^{n_2}-1)d_1(x)},  & \mbox{ if } \Lambda(\theta)=0 \\
            \frac{(x^n-1)(x-1)}{(x^{n_1}-1)(x^{n_2}-1)d_0(x)}, & \mbox{ if } \Lambda(\theta)=1.  
                                               \end{array} 
                                    \right.                    
           \end{eqnarray*} 
           In this case, the cyclic code $\calC_\lambda$ over 
          $\gf(q)$ defined by the sequence $\lambda^{\infty}$ has generator polynomial $\m_\lambda(x)$  
          above and parameters $\left[n, k, d \right]$, where  $k=\frac{(n_1+1)(n_2+1)-2}{2}$ and 
          $d$ has the lower bound of Theorem \ref{thm-2Pbound153}.            
           
\item If ${n_1} \equiv -1 \pmod 8$ and ${n_2} \equiv -1 \pmod 8$ or ${n_1} \equiv 
           3 \pmod 8$ and ${n_2} \equiv 3 \pmod 8$, we have  
           \begin{eqnarray*} 
&&             \ls_\lambda=n-\frac{(n_1+1)(n_2-1)+2}{2},              \\ 
&&             \m_\lambda(x)=\left\{ \begin{array}{ll} 
             \frac{x^n-1}{(x^{n_2}-1)d_0(x)},  & \mbox{ if } \Lambda(\theta)=0 \\
             \frac{x^n-1}{(x^{n_2}-1)d_1(x)},  & \mbox{ if } \Lambda(\theta)=1.  
                                               \end{array} 
                                    \right.                    
           \end{eqnarray*} 
           In this case, the cyclic code $\calC_\lambda$ over 
          $\gf(q)$ defined by the sequence $\lambda^{\infty}$ has generator polynomial $\m_\lambda(x)$  
          above and parameters $\left[n, k, d \right]$, where  $k=\frac{(n_1+1)(n_2-1)+2}{2}$ 
          and 
          $d$ has the lower bound of Theorem \ref{thm-2Pbound154}.                       
           
\item If ${n_1} \equiv 1 \pmod 8$ and ${n_2} \equiv 1 \pmod 8$ or ${n_1} \equiv 
           -3 \pmod 8$ and ${n_2} \equiv -3 \pmod 8$, we have  
           \begin{eqnarray*} 
&&            \ls_\lambda=n-\frac{(n_1-1)(n_2+1)+2}{2},              \\   
&&           \m_\lambda(x)=\left\{ \begin{array}{ll} 
                          \frac{x^n-1}{(x^{n_1}-1)d_0(x)}, & \mbox{ if } \Lambda(\theta)=0 \\
                           \frac{x^n-1}{(x^{n_1}-1)d_1(x)},  & \mbox{ if } \Lambda(\theta)=1.  
                                               \end{array} 
                                    \right.                    
           \end{eqnarray*}  
           In this case, the cyclic code $\calC_\lambda$ over 
          $\gf(q)$ defined by the sequence $\lambda^{\infty}$ has generator polynomial $\m_\lambda(x)$  
          above and parameters $\left[n, k, d \right]$, where  $k=\frac{(n_1-1)(n_2+1)+2}{2}$ 
          and 
          $d$ has the lower bound of Theorem \ref{thm-2Pbound154}.             
                           
\end{enumerate} 
\end{corollary} 

\begin{example}\label{exam-2p90009} 
Let $(p, m, n_1, n_2)=(2,1,3,5)$.  Then $q=2$, $n=15$, and 
$\calC_\lambda$ is a $[15,11,3]$ cyclic code over $\gf(q)$ 
with generator polynomial 
$  
x^4 + x^3 + 1.
$
This code is optimal.  
\end{example} 

\begin{remark} 
It was proved in \cite{Whit62} that $C_1=N_1 \cup D_1^{(2)}$ is a difference set when 
$n_2=n_1+2$ and $n_1$ are primes. Example \ref{exam-2p90009} shows that a 
difference set may give an optimal cyclic code. 
\end{remark}

\begin{example}\label{exam-2pOct15} 
Let $(p, m, n_1, n_2)=(2,1,5,3)$.  Then $q=2$, $n=15$, and 
$\calC_\lambda$ is a $[15,5,7]$ cyclic code over $\gf(q)$ 
with generator polynomial
$ 
x^{10} + x^9 + x^8 + x^6 + x^5 + x^2 + 1. 
$ 
This code is optimal. 
\end{example} 

\begin{example}\label{exam-2p151} 
Let $(p, m, n_1, n_2)=(2,1,3,7)$.  Then $q=2$, $n=21$, and  
$\calC_\lambda$ is a $[21,7,3]$ cyclic code over $\gf(q)$ 
with generator polynomial $x^{14}+x^7+1$. This is a bad 
cyclic code due to its poor minimum distance. The code in this 
case is bad because $q \not\in D_0^{(2)}$. 
\end{example} 

\begin{remark} 
It was proved in \cite{Ding98,German} that $C_1=N_1 \cup D_1^{(2)}$ is an almost difference set when 
$n_2=n_1+4$ and $n_1$ both are primes. Example \ref{exam-2p151} shows that this difference set 
gives a bad cyclic code over $\gf(2)$. Later we will see that it may give an almost optimal cyclic 
code over other fields $\gf(q)$. 
\end{remark}

\begin{example} 
Let $(p, m, n_1, n_2)=(2,1,7,5)$.  Then $q=2$, $n=35$, and 
$\calC_\lambda$ is a $[35,11,5]$ cyclic code over $\gf(q)$ 
with generator polynomial
\begin{eqnarray*}
x^{24} + x^{23} + x^{19} + x^{18} + x^{17} + x^{16} + x^{14} + x^{13} +\\ 
 x^{12} + x^{11} +
    x^{10} + x^8 + x^7 + x^6 + x^5 + x + 1. 
\end{eqnarray*}
\end{example}

\subsubsection{The case $q=3$}

The following corollary follows directly from  Theorem \ref{thm-mmm}. 

\begin{corollary}\label{cor-lsq=3} 
Let $q=3$. We have the following conclusions: 
\begin{enumerate}
\item If ${n_1} \equiv 1 \pmod{12}$ and ${n_2} \equiv -1 \pmod{12} $ or ${n_1} \equiv 
           -5 \pmod{12}$ and ${n_2} \equiv 5 \pmod{12}$, we have  
           \begin{eqnarray*} 
&&            \ls_{\lambda}=n-\frac{(n_1-1)(n_2-1)}{2}, \\   
&&            \m_\lambda(x)=\left\{ \begin{array}{ll} 
                                               \frac{x^{{n}}-1}{d_0(x)}, & \mbox{ if } \Lambda(\theta)=0 \\
                                               \frac{x^{{n}}-1}{d_1(x)}, & \mbox{ if } \Lambda(\theta)=-1.  
                                               \end{array} 
                                    \right.                                                            
           \end{eqnarray*}    
           In this case, the cyclic code $\calC_\lambda$ over 
          $\gf(q)$ defined by the sequence $\lambda^{\infty}$ has generator polynomial $\m_\lambda(x)$  
          above and parameters $\left[n, k, d \right]$, where $k=\frac{(n_1-1)(n_2-1)}{2}$.                           

\item If ${n_1} \equiv -1 \pmod{12}$ and ${n_2} \equiv 5 \pmod{12}$ or ${n_1} \equiv 
           5 \pmod{12}$ and ${n_2} \equiv -1 \pmod{12}$, we have  
           \begin{eqnarray*} 
            \ls_{\lambda}=n-n_2, \ \ \m_\lambda(x)=\frac{x^{{n}}-1}{x^{n_2}-1}. 
           \end{eqnarray*} 
           In this case, the cyclic code $\calC_\lambda$ over 
          $\gf(q)$ defined by the sequence $\lambda^{\infty}$ has generator polynomial $\m_\lambda(x)$  
          above and parameters $\left[n, n_2, n_1 \right]$, where the minimum weight follows 
          from Theorem \ref{thm-2Pbound151}.                           
           
\item If ${n_1} \equiv -5 \pmod{12}$ and ${n_2} \equiv 1 \pmod{12}$ or ${n_1} \equiv 
           1 \pmod{12}$ and ${n_2} \equiv -5 \pmod{12}$, we have  
           \begin{eqnarray*} 
            \ls_{\lambda}=n-n_1, \ \ \m_\lambda(x)=\frac{x^{{n}}-1}{x^{n_1}-1}. 
           \end{eqnarray*}  
           In this case, the cyclic code $\calC_\lambda$ over 
          $\gf(q)$ defined by the sequence $\lambda^{\infty}$ has generator polynomial $\m_\lambda(x)$  
          above and parameters $\left[n, n_1, n_2 \right]$, where the minimum weight follows 
          from Theorem \ref{thm-2Pbound151}.                         
                     
\item If ${n_1} \equiv -1 \pmod{12}$ and ${n_2} \equiv 1 \pmod{12}$ or ${n_1} \equiv 
           5 \pmod{12}$ and ${n_2} \equiv -5 \pmod{12}$, we have  
           \begin{eqnarray*} 
&&             \ls_{\lambda}=n-\frac{(n_1+1)(n_2+1)-2}{2},  \\  
&&            \m_\lambda(x)=\left\{ \begin{array}{ll} 
              \frac{(x^n-1)(x-1)}{(x^{n_1}-1)(x^{n_2}-1)d_0(x)}, & \mbox{ if } \Lambda(\theta)=0 \\
              \frac{(x^n-1)(x-1)}{(x^{n_1}-1)(x^{n_2}-1)d_1(x)}, & \mbox{ if } \Lambda(\theta)=-1.  
                                               \end{array} 
                                    \right.                    
           \end{eqnarray*}  
           In this case, the cyclic code $\calC_\lambda$ over 
          $\gf(q)$ defined by the sequence $\lambda^{\infty}$ has generator polynomial $\m_\lambda(x)$  
          above and parameters $\left[n, k, d \right]$, where  $k=\frac{(n_1+1)(n_2+1)-2}{2}$ 
          and 
          $d$ has the lower bound of Theorem \ref{thm-2Pbound153}.            

\item If ${n_1} \equiv 1 \pmod{12}$ and ${n_2} \equiv 5 \pmod{12}$ or ${n_1} \equiv 
           -5 \pmod{12}$ and ${n_2} \equiv -1 \pmod{12}$, we have  
           \begin{eqnarray*} 
            \ls_{\lambda}={n}, \ \ \m_\lambda(x)=x^{{n}}-1. 
           \end{eqnarray*}     
           In this case, the cyclic code $\calC_\lambda$ over 
          $\gf(q)$ defined by the sequence $\lambda^{\infty}$ has generator polynomial $\m_\lambda(x)$  
          above and parameters $\left[n, 0, 0 \right]$.                   

\item If ${n_1} \equiv -1 \pmod{12}$ and ${n_2} \equiv -5 \pmod{12}$ or ${n_1} \equiv 
           5 \pmod{12}$ and ${n_2} \equiv 1 \pmod{12}$, we have  
           \begin{eqnarray*} 
            \ls_{\lambda}={n}-({n_1}+{n_2}-1), \ \ \m_\lambda(x)=\frac{(x^{{n}}-1)(x-1)}{(x^{n_1}-1)(x^{n_2}-1)}. 
           \end{eqnarray*}       
           In this case, the cyclic code $\calC_\lambda$ over 
          $\gf(q)$ defined by the sequence $\lambda^{\infty}$ has generator polynomial $\m_\lambda(x)$  
          above and parameters $\left[n, n_1+n_2-1, d \right]$, where $d$ has the lower bound 
          of Theorem \ref{thm-2Pbound152}.                          
           
\item If ${n_1} \equiv 5 \pmod{12}$ and ${n_2} \equiv 5 \pmod{12}$ or ${n_1} \equiv 
           -1 \pmod{12}$ and ${n_2} \equiv -1 \pmod{12}$, we have  
           \begin{eqnarray*} 
&&             \ls_{\lambda}=n-\frac{(n_1+1)(n_2-1)+2}{2},              \\  
&&            \m_\lambda(x)=\left\{ \begin{array}{ll} 
               \frac{x^n-1}{(x^{n_2}-1)d_0(x)},   & \mbox{ if } \Lambda(\theta)=0 \\
               \frac{x^n-1}{(x^{n_2}-1)d_1(x)},   & \mbox{ if } \Lambda(\theta)=-1.  
                                               \end{array} 
                                    \right.                    
           \end{eqnarray*} 
           In this case, the cyclic code $\calC_\lambda$ over 
          $\gf(q)$ defined by the sequence $\lambda^{\infty}$ has generator polynomial $\m_\lambda(x)$  
          above and parameters $\left[n, k, d \right]$, where  $k=\frac{(n_1+1)(n_2-1)+2}{2}$ 
          and 
          $d$ has the lower bound of Theorem \ref{thm-2Pbound154}.            
           
\item If ${n_1} \equiv 1 \pmod{12}$ and ${n_2} \equiv 1 \pmod{12}$ or ${n_1} \equiv 
           -5 \pmod{12}$ and ${n_2} \equiv -5 \pmod{12}$, we have  
           \begin{eqnarray*} 
&&             \ls_{\lambda}=n-\frac{(n_1-1)(n_2+1)+2}{2},              \\  
&&           \m_\lambda(x)=\left\{ \begin{array}{ll} 
               \frac{x^n-1}{(x^{n_1}-1)d_0(x)},   & \mbox{ if } \Lambda(\theta)=0 \\
               \frac{x^n-1}{(x^{n_1}-1)d_1(x)},   & \mbox{ if } \Lambda(\theta)=-1.  
                                               \end{array} 
                                    \right.                    
           \end{eqnarray*}            
           In this case, the cyclic code $\calC_\lambda$ over 
          $\gf(q)$ defined by the sequence $\lambda^{\infty}$ has generator polynomial $\m_\lambda(x)$  
          above and parameters $\left[n, k, d \right]$, where  $k=\frac{(n_1-1)(n_2+1)+2}{2}$ 
          and 
          $d$ has the lower bound of Theorem \ref{thm-2Pbound154}.            
                             
\end{enumerate} 
\end{corollary} 

\begin{example} 
Let $(p, m, n_1, n_2)=(3,1,7,5)$.  Then $q=3$, $n=35$, and  
$\calC_\lambda$ is a $[35,12,12]$ cyclic code over $\gf(q)$ 
with generator polynomial
\begin{eqnarray*}
x^{23} + 2x^{21} + 2x^{19} + 2x^{18} + x^{16} + 2x^{14} + x^{13} + 2x^{12} + \\ 
    2x^{11} + 2x^{10} + x^8 + 2x^6 + 2x^5 + 2x^4 + x^3 + x^2 + 2x + 2. 
\end{eqnarray*} 
The best ternary linear code known of length 35 and dimension 12 has minimum distance 14.  
\end{example} 

\begin{example}\label{exam-2p152} 
Let $(p, m, n_1, n_2)=(3,1,5,7)$.  Then $q=3$, $n=35$, and 
 $\calC_\lambda$ is a $[35,23,5]$ cyclic code over $\gf(q)$ 
with generator polynomial
$
x^{12} + 2x^{11} + 2x^{10} + x^9 + 2x^8 + x^7 + x^5 + 2x^4 + x^2 + 1. 
$  
The best ternary linear code known of length 35 and dimension 23 has minimum distance 6.  
\end{example} 

\begin{remark} 
It was proved in \cite{Whit62} that $C_1=N_1 \cup D_1^{(2)}$ is a difference set when 
$n_2=n_1+2$ and $n_1$ are primes. Example \ref{exam-2p152} demonstrates that 
a difference set may give a very good cyclic code. 
\end{remark}


Finally, we present results on the minimum distance of some of the cyclic codes 
$\C_{\lambda}$.

\begin{theorem}\label{thm-2Pbound151}
The cyclic code over $\gf(q)$ with the generator polynomial  $g(x)=(x^n-1)/(x^{n_i}-1)$ 
has parameters  $[n, n_i, d_i]$, where   
\begin{eqnarray}\label{eqn-2Pbound151} 
d_i = n_{i-(-1)^i},   
\end{eqnarray}
where $i \in \{1,2\}$. 
\end{theorem} 

\begin{example} 
Let $q=2$ and $(n_1, n_2)=(7,5)$. Then the cyclic code over $\gf(q)$ with the generator polynomial  
$g(x)=(x^n-1)/(x^{n_1}-1)$ has parameters  $[35, 7, 5]$.   
\end{example} 

\begin{theorem}\label{thm-2Pbound152}
Let $\C_{(n_1, n_2, q)}$ denote the cyclic code over $\gf(q)$ with the generator polynomial 
$$g(x)=(x-1)(x^n-1)/(x^{n_1}-1)(x^{n_2}-1).$$ Then the code $\C_{(n_1, n_2, q)}$  has parameters  
$[n, n_1+n_2-1, d_{(n_1,n_2)}]$, where   
\begin{eqnarray}\label{eqn-2Pbound152}
d_{(n_1, n_2)}  = \min(n_1, n_2). 
\end{eqnarray}
\end{theorem} 

\begin{example} 
Let $q=2$ and $(n_1, n_2)=(7,5)$. The cyclic code over $\gf(q)$ with the generator polynomial  
$g(x)=(x^n-1)/(x^{n_1}-1)(x^{n_2}-1)$ has parameters  $[35, 11, 5]$.  
\end{example}

\begin{theorem}\label{thm-2Pbound153}
Assume that $q \in D_0^{(2)}$. Let $\C_{(i, q)}$ denote the cyclic code 
over $\gf(q)$ with the generator polynomial $d_i(x)$ for $i=0$ and $i=1$. 
Then the code $\C_{(i, q)}$  has parameters  
$[n, ((n_1+1)(n_2+1)-2)/2, d_i]$, where   
\begin{eqnarray}\label{eqn-2Pbound153a}
d_i \ge \left\lceil \sqrt{ \min(n_1, n_2)} \right\rceil.  
\end{eqnarray}
If $-1 \in D_1^{(2)}$, we have  
\begin{eqnarray}\label{eqn-2Pbound153b}
d_i^2-d_i+1 \ge \min(n_1, n_2).  
\end{eqnarray}
\end{theorem}

\begin{example} 
Let $(p, m, n_1, n_2)=(3,1,5,7)$.  Then $q=3$, $n=35$, and $\calC_{(0, q)}$ is a 
$[35,23,5]$ cyclic code over $\gf(q)$ 
with generator polynomial
$ 
x^{12} + 2x^{11} + 2x^{10} + x^9 + 2x^8 + x^7 + x^5 + 2x^4 + x^2 + 1. 
$ 
In this case, $-1 \in D_1^{(2)}$ and $d_{(n_1, n_2)}=5$. We have  
$$ 
d_0^2-d_0+1 \ge 5. 
$$ 
Hence the lower bounds of (\ref{eqn-2Pbound153a}) and (\ref{eqn-2Pbound153b}) are 3. 
In this case, the two lower bounds are not met.   
\end{example} 

\begin{theorem}\label{thm-2Pbound154} 
Let $q \in D_0^{(2)}$. 
Let $\C_{(n_1, n_2, q)}^{(i,j)}$ denote the cyclic code over $\gf(q)$ with the generator polynomial 
$$
g^{(i,j)}(x)=\frac{(x^{n_i}-1)}{x-1} d_j(x), 
$$ 
and let $d_{(n_1,n_2,q)}^{(i,j)}$ denote the minimum 
distance of this code, where $i \in \{1, 2\}$ and $j \in \{0, 1\}$. 
Then the code $\C_{(n_1, n_2, q)}^{(i,j)}$  has parameters  
$$\left[n, \frac{(n_i-1)(n_{i-(-1)^i}+1)+2}{2}, d_{(n_1,n_2,q)}^{(i,j)}\right],$$ 
where   
\begin{eqnarray}\label{eqn-2Pbound154a}
d_{(n_1,n_2,q)}^{(i,j)} \ge \left\lceil \sqrt{n_{i}} \right\rceil.  
\end{eqnarray}
If $-1 \in D_1^{(2)}$, we have  
\begin{eqnarray}\label{eqn-2Pbound155b}
\left(d_{(n_1,n_2,q)}^{(i,j)}\right)^2 -  d_{(n_1,n_2,q)}^{(i,j)} + 1 \ge n_{i}. 
\end{eqnarray}
\end{theorem} 

\begin{example} 
Let $(p, m, n_1, n_2)=(2,1,3,11)$.  Then $q=2$, $n=33$, and  
$\calC_\lambda$ is a $[33,21,3]$ cyclic code over $\gf(q)$ 
with generator polynomial
\begin{eqnarray*}
x^{12} + x^9 + x^7 + x^6 + x^5 + x^3 + 1= \frac{x^{n_1}-1}{x-1} d_0(x). 
\end{eqnarray*} 
In this case, the lower bound of (\ref{eqn-2Pbound154a}) is 2, while the actual minimum 
distance is 3. 
\end{example}

\section{Concluding remarks} 

In addition to the results presented in this paper, cyclic codes with interesting parameters were constructed with 
cyclotomic sequences of order $4$ in \cite{Ding13,Wang17}.   

Recall that every cyclic code over any finite field could be expressed as a code $\C_s$ for a sequence $s^\infty$. 
This approach can produce all cyclic codes over finite fields, including BCH codes. It is thus no surprise 
that some of the codes from Dickson polynomials are in fact BCH codes. Since the sequence approach is fundamental, 
it produces both good and bad cyclic codes. It is open what sequences over a finite field give cyclic codes   
with optimal parameters. There are many open problems in this direction. 

Though a considerable amount of progress on this approach of constructing cyclic codes with sequences has been made, a lot of investigation should be further done, as there is a huge number of constructions of sequences in the literature. The reader is cordially invited to join the journey in this direction. 

\section*{Acknowledgements} 
The author acknowledges the support from the Hong Kong Grants Council, under Proj. No. $16301020$.

\end{document}